\documentclass[apj]{emulateapj}
\usepackage{lscape} % for rotating tables
\newcommand{\etal}{et~al.}

\newcommand{\cgsflux}{erg~s$^{-1}$~cm$^{-2}$}

\newcommand{\OIIlong}{{\rm O}\kern 0.1em{\sc ii}~$\lambda 3727$} 
\newcommand{\OII}{{\rm O}\kern 0.1em{\sc ii}} 
\newcommand{\Lsun}{L$_{\odot}$}
\newcommand{\Msun}{M$_{\odot}$}

\newcommand{\nufnu}{$\nu f_{\nu}$}
\newcommand{\fnu}{$f_{\nu}$}
\newcommand{\microJy}{$\mu$Jy}
\newcommand{\imageA}{SMM J163555.2+661238}
\newcommand{\imageB}{SMM J163554.2+661225}

\slugcomment{Resumbitted, 6 Sept 2007}

\shorttitle{IRS spectra of lensed galaxies}
\shortauthors{Rigby \etal}

\begin{document}

\title{Mid-Infrared Spectroscopy of Lensed Galaxies at $1<z<3$:  
The Nature of Sources Near the MIPS Confusion Limit}
\author{J.~R.~Rigby\altaffilmark{1,2,3},
D.~Marcillac\altaffilmark{1},  
E.~Egami\altaffilmark{1}, 
G.~H.~Rieke\altaffilmark{1},
J.~Richard\altaffilmark{4},
J.-P.~Kneib\altaffilmark{5},
D.~Fadda\altaffilmark{6}, 
C.~N.~A.~Willmer\altaffilmark{1}, 
C.~Borys\altaffilmark{6},
P.~P.~ van der Werf\altaffilmark{6}, 
P.~G.~P\'erez-Gonz\'alez\altaffilmark{7},
K.~K.~Knudsen\altaffilmark{8}, and
C.~Papovich\altaffilmark{1}}

\altaffiltext{1}{Steward Observatory, University of Arizona, 933 N. 
Cherry Ave., Tucson, AZ 85721}
\altaffiltext{2}{Current address:  
Observatories, Carnegie Institution of Washington,
813 Santa Barbara St., Pasadena, CA 91101}
\altaffiltext{3}{Spitzer Fellow}
\altaffiltext{4}{Department of Astronomy, Caltech, 
1200 E. California Blvd, Pasadena, CA 91125}
\altaffiltext{5}{Laboratoire d'Astrophysique de Marseille}
\altaffiltext{6}{Leiden Observatory, Leiden University, P. O. 
Box 9513, NL-2300 RA Leiden, The Netherlands}
\altaffiltext{7}{Departamento de Astrof\'isica y CC. de la Atm\'osfera, 
Facultad de CC. F\'isicas, Universidad Complutense de Madrid, 
28040 Madrid, Spain}
\altaffiltext{8}{Max-Planck-Institut f\"ur Astronomie, 
K\"onigstuhl 17, D-69117 Heidelberg, Germany}
\email{jrigby@ociw.edu}

\begin{abstract}

We present Spitzer/IRS mid-infrared spectra for 15 gravitationally lensed, 
24~\micron--selected galaxies, and combine the results with 4 additional
very faint galaxies with IRS spectra in the literature.
The median intrinsic 24~\micron\ flux density of the sample is 130~\microJy,
enabling a systematic survey of the spectral properties of the very faint
24~\micron\ sources that dominate the number counts of Spitzer 
cosmological surveys. 
Six of the 19 galaxy spectra ($32\%$)  show the strong mid-IR continuua 
expected of AGN; X-ray detections confirm the presence of AGN
in three of these cases, and reveal AGNs in two other galaxies.
These results suggest that nuclear accretion may contribute more flux 
to faint 24~\micron--selected samples than previously assumed.
Almost all the spectra show some aromatic (PAH) emission features; 
the measured aromatic flux ratios do not show evolution from $z=0$.
In particular, the  high S/N mid-IR spectrum of \imageB\
agrees remarkably well with low--redshift, lower--luminosity templates.
%neither it nor the full sample show evidence that the
%flux ratios of the aromatic features in star--forming galaxies 
%have evolved with redshift.
We compare the rest-frame 8~\micron\ and total infrared luminosities
of star--forming galaxies,
and find that the behavior of this ratio with total IR luminosity
has evolved modestly from z$=$2 to z$=$0.  
Since the high aromatic--to--continuum flux ratios in these galaxies rule out
a dominant contribution by AGN, this finding implies systematic evolution in
the structure and/or metallicity of infrared sources with redshift.  
It also has implications for the estimates of star forming
rates inferred from 24~\micron\ measurements, in the sense that at 
$z \sim 2$, a given observed frame 24~\micron\ luminosity corresponds to a
lower bolometric luminosity than would be inferred from low-redshift
templates of similar luminosity at the corresponding rest wavelength.
\end{abstract}

\keywords{galaxies---infrared: galaxies: high-redshift---galaxies: evolution}

\section{Introduction}
The Spitzer Space Telescope \citep{werner-spitz} 
has been tremendously successful at
detecting star--forming galaxies and active galactic nuclei (AGN) at
$z>1$ by their emission in the observed--frame 24~\micron\ band.
Down to the confusion limit of $\sim 50$~$\mu$Jy at 24~\micron\
\citep{dole04},
the MIPS instrument \citep{rieke-mips} can detect
luminous infrared galaxies (LIRGs)\footnote{Defined as $11 < \log L(TIR)$ \Lsun $< 12$,
where L(TIR) is the total infrared luminosity between 8 and 1000~\micron.}
out to $z \sim 2$, 
ultra-luminous infrared galaxies (ULIRGs)\footnote{Defined as 
$12 < \log L(TIR)< 13$~\Lsun.} out to $z\sim3$,
and hyper-luminous infrared galaxies (HLIRGs)\footnote{Defined as
$\log L(TIR)> 13$~\Lsun.} out to even higher redshifts.
 
Detectability at $z>0.7$ is increased by 
strong aromatic emission features, at rest--frame 6--12~\micron,
that pass through the 24~\micron\ band.  Indeed, photometric 
redshifts place $\sim 25$--$30\%$ of the faint 24~\micron\ sources
at redshifts above 1.4 \citep{emeric05,pablo05,caputi06,wang06}.

For extremely high--luminosity galaxies, spectra have been obtained
with the Infrared Spectrograph (IRS, \citealt{houck-irs}),
e.g.~\citet{houck-highz,yan07,md07}.  
However, since spectral properties depend strongly on luminosity, 
results obtained for hyper-luminous galaxies may have limited 
applicability for the bulk of the IR--detected population.
Spectra of lower--luminosity galaxies have been obtained only in a few
cases (e.g. \citealt{lutz05,teplitz07}).

Numerous authors (e.g. \citealt{caputi06,choi06,reddy06}) 
have used the observed 24~\micron\ band
as an estimator of star formation rates.  Since most of the 
infrared power is radiated 
redward of the observed band, the 24~\micron\ diagnostic must 
be calibrated using low--redshift templates
(e.g.~\citealt{charyelbaz, dalehelou,lagache04,brandl06,armus07}).
The central assumption in these works is that the spectra 
and spectral energy distributions of high-z galaxies 
are matched accurately by the low--redshift templates.
For $0.7<z<3$, the 24~\micron\ band fluxes  are strongly
influenced by aromatic band  emission.  Evolution over
time either in the behavior of these aromatic bands, in metallicity,
or in the geometry and radiative transfer within these infrared 
sources, could all undermine this central assumption.

To probe these possibilities, we have been obtaining IRS spectra 
for intrinsically faint 24~\micron--selected galaxies at $1\la z \la 3$.
Our targets are strongly lensed by the gravitational
potential of foreground clusters of galaxies,  such that their fluxes
are amplified by factors of 3--25.  
Though these galaxies have observed 24~\micron\ flux densities of
$\sim 1$~mJy, sufficient for useful, efficient spectroscopy with IRS, 
their intrinsic luminosities are down in the LIRG to ULIRG range.

In this paper, we present IRS spectra for the first
15 galaxies in the sample.    The median observed flux density is 
f$_{\nu}$(24~\micron) $=  0.82$~mJy, but the median 
intrinsic flux density is nearly an order of magnitude lower.
We compare their aromatic feature
flux ratios, and the ratios of their X-ray, aromatic, and total 
infrared luminosities to those of low--redshift LIRGs and ULIRG samples.
We examine in detail the high--quality spectrum of \imageB,
%SMM J163554.2+661225, 
a LIRG at $z=2.516$.
We merge our sample with four additional objects from the literature
\citep{lutz05,teplitz07} to provide a first systematic look at the 
IRS--measured properties of the faint 24~\micron\ galaxy population.

We assume $\Omega_m = 0.265$, $\Omega_{\Lambda} = 0.735$, and
H$_0 = 72$~km~s$^{-1}$~Mpc$^{-1}$ \citep{spergelwmap1,freedman01}.

\section{Sample Selection}
Our goal was to select sources with intrinsic flux densities close to,
or even below, the MIPS 24~\micron\  confusion limit.  
We first obtained deep 24~\micron\ images of galaxy clusters,
generally identified as being massive through high X-ray luminosity.
From these 24~\micron\ images, we selected sources within 1.5\arcmin\
of the cluster center that:
\begin{itemize}
\item had observed 24~\micron\ flux densities above 0.4~mJy 
(to make efficient use of IRS).
\item had faint, irregular, or arc--like HST counterparts.
This excluded galaxies at the cluster redshift.\footnote{A2219a
and A2261a lack HST imagery; we instead used ground-based F606W images from
the Steward Observatory 90 inch telescope.  Both galaxies are known 
sub-mm sources (see Table~\ref{tab:sample}), and can thus be added
to the sample despite the lack of high--resolution optical images.}
\item had known spectroscopic or probable photometric redshifts $\ga 1.0$. 
\item had lensing amplification factors calculated to be greater than 3.  
This restricted selection to clusters with good lensing models.
Since amplification depends on the source redshift, 
if the redshift was unknown, we calculated 
the amplification for $1<z<3$ and required amplifications factors 
greater than 3 over that range.  
\end{itemize}

\begin{figure}
\figurenum{1}
\includegraphics[height=1.5in]{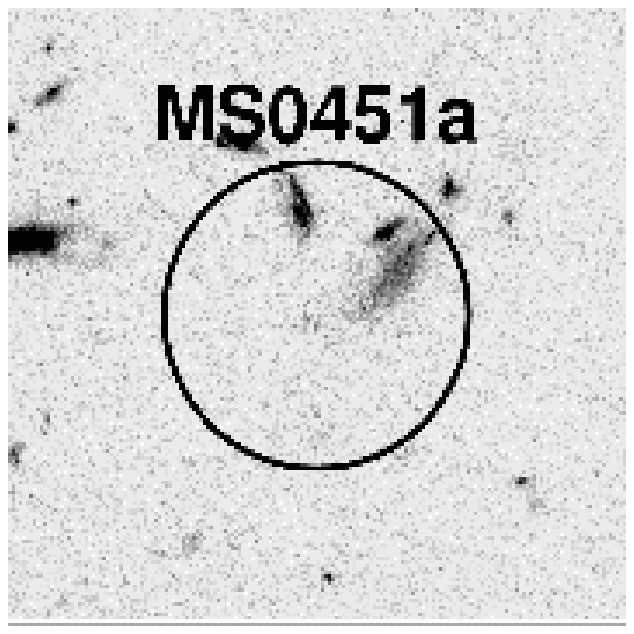} %{MS0451a-s.ps}
\includegraphics[height=1.5in]{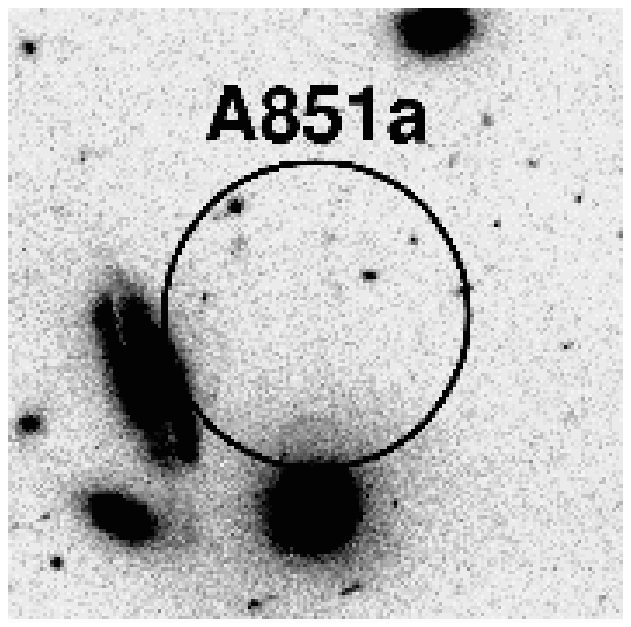} %{A851a-s.ps}
\includegraphics[height=1.5in]{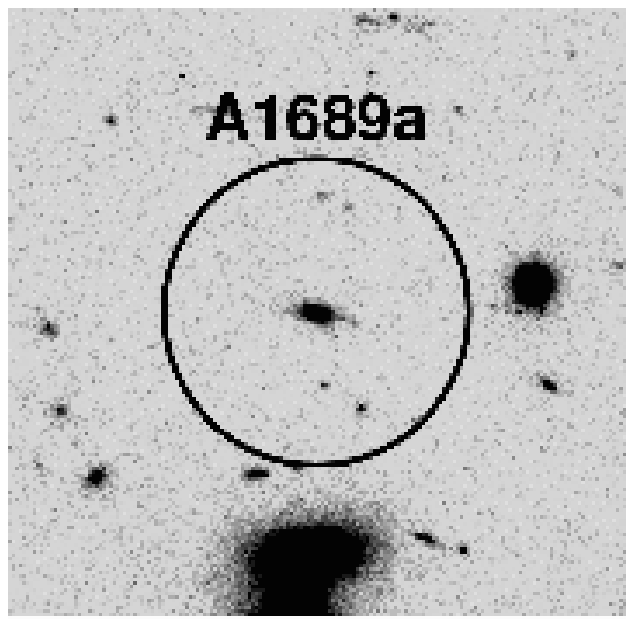} %{A1689a-s.ps}
\includegraphics[height=1.5in]{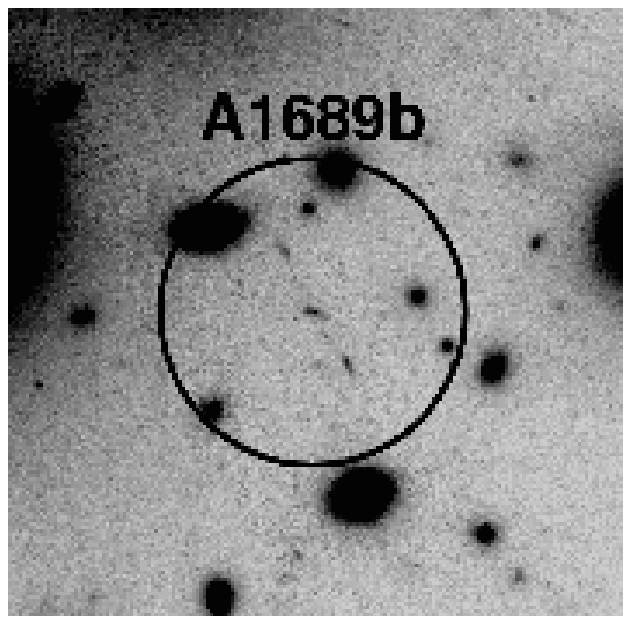} %{A1689b-s.ps}
\includegraphics[height=1.5in]{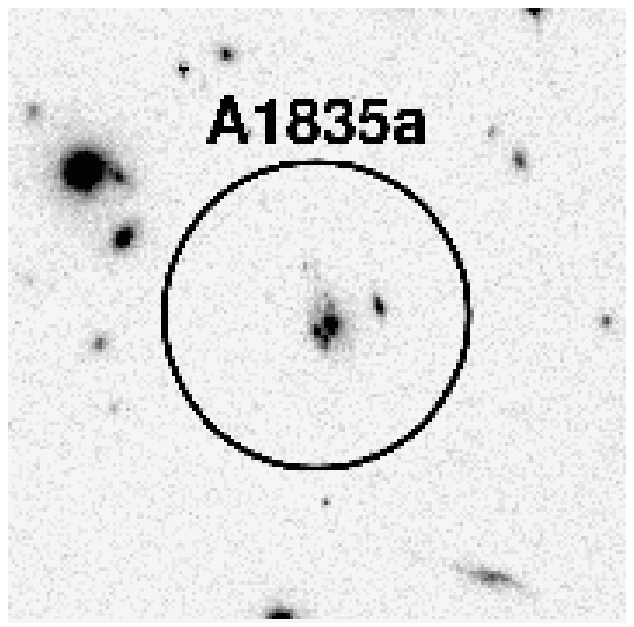} %{A1835a-s.ps}
\includegraphics[height=1.5in]{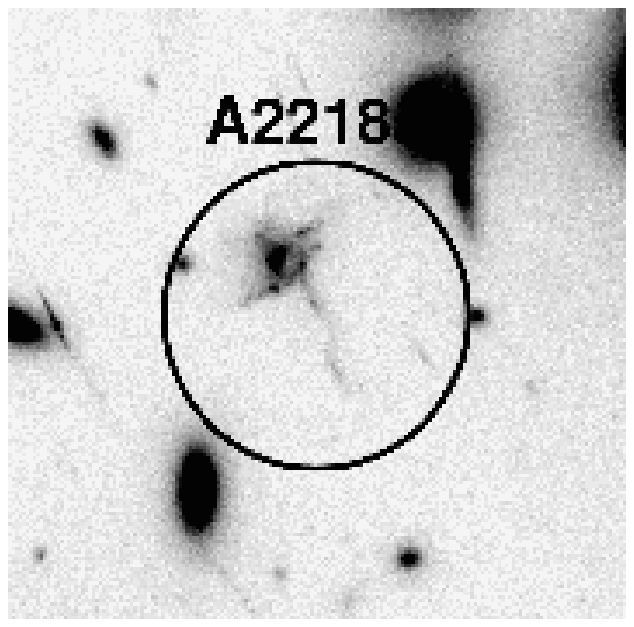} %{A2218a-s.ps}
\includegraphics[height=1.5in]{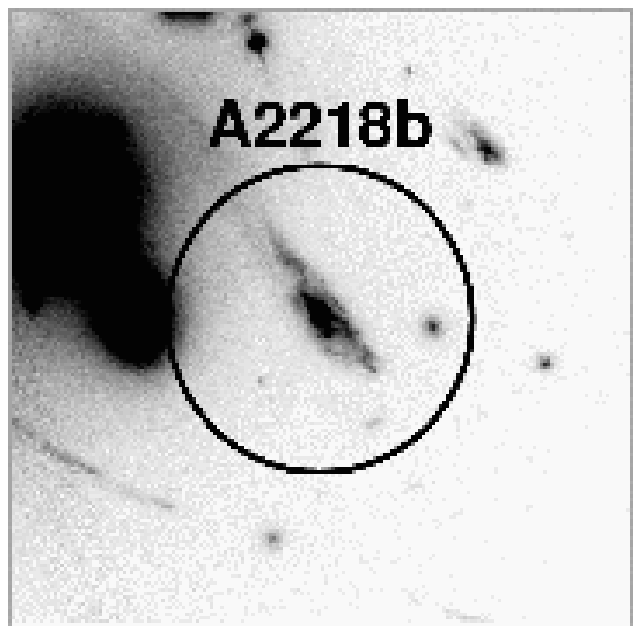} %{A2218b-s.ps}
\includegraphics[height=1.5in]{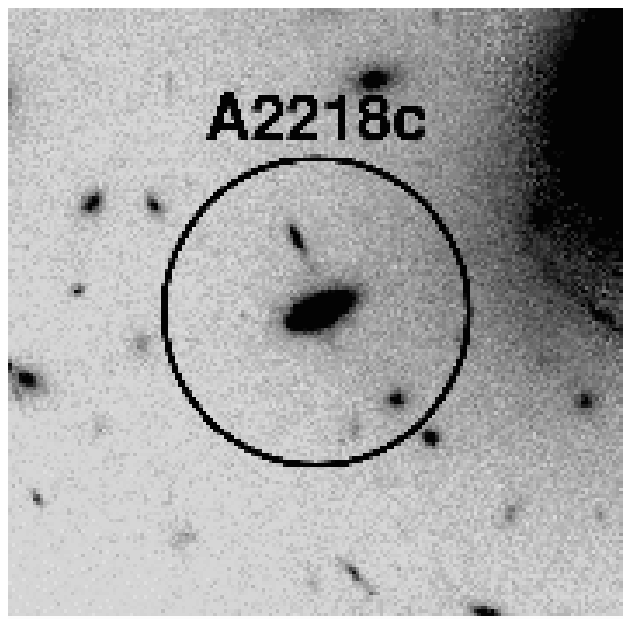} %{A2218c-s.ps}
\includegraphics[height=1.5in]{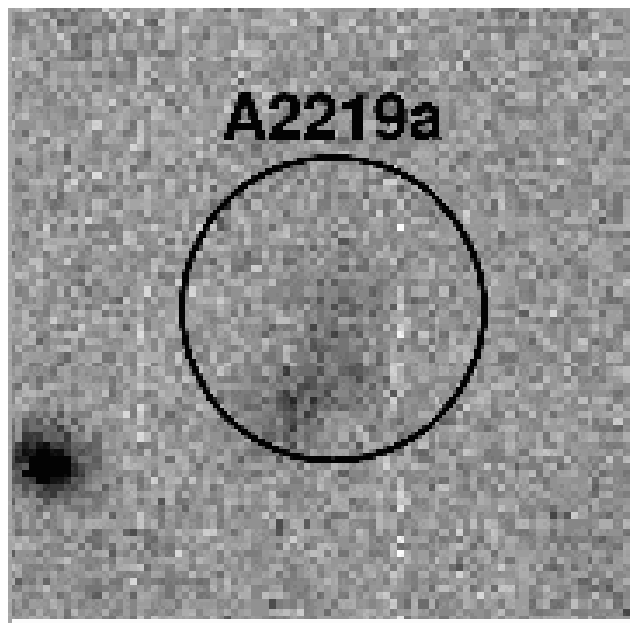} %{A2219a-s.ps}
\includegraphics[height=1.5in]{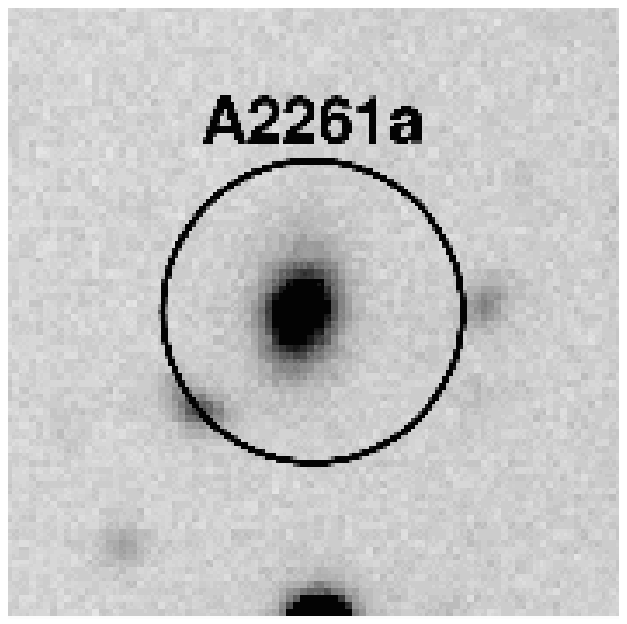} %{A2261a-s.ps}
\includegraphics[height=1.5in]{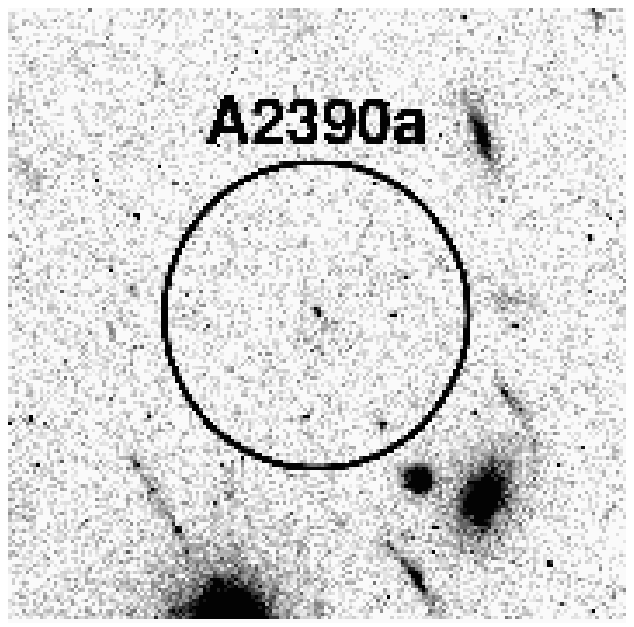} %{A2390a-s.ps}
\includegraphics[height=1.5in]{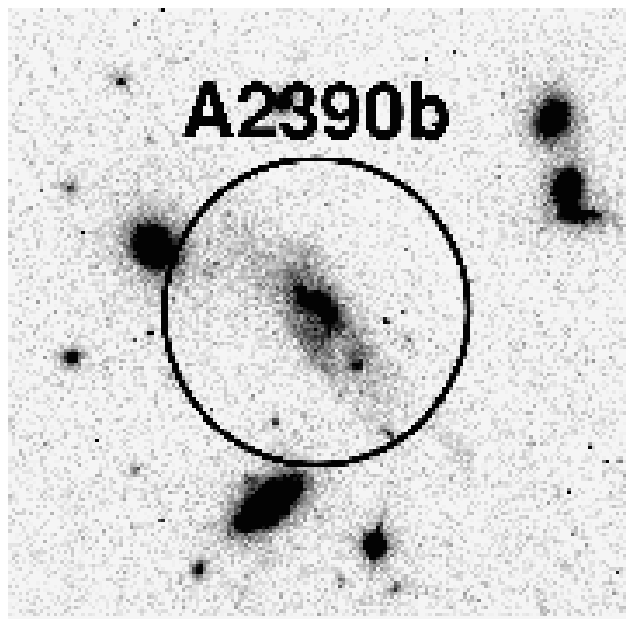} %{A2390b-s.ps}
\includegraphics[height=1.5in]{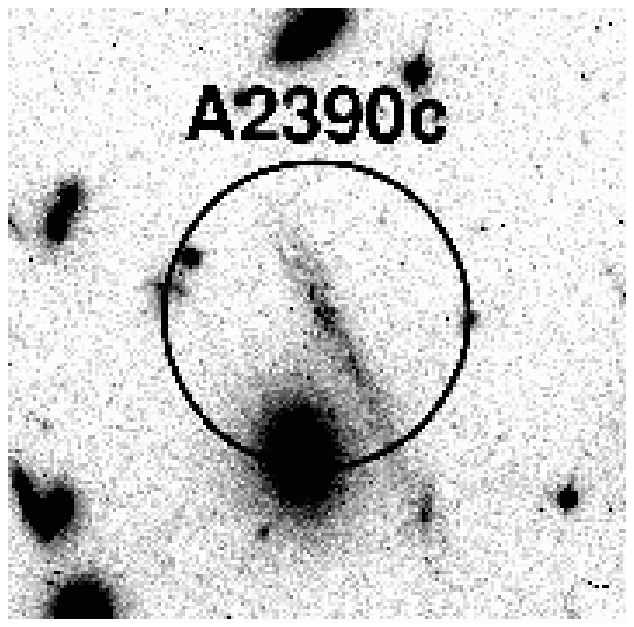} %{A2390c-s.ps}
\includegraphics[height=1.5in]{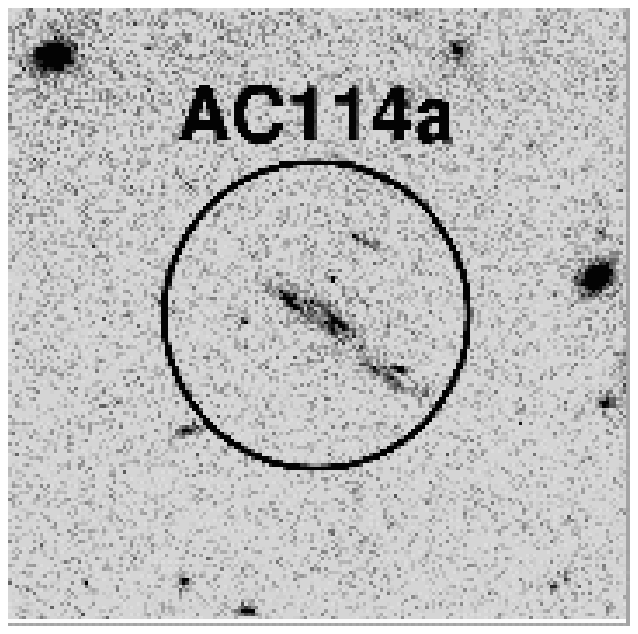} %{AC114a-s.ps}
\includegraphics[height=1.5in]{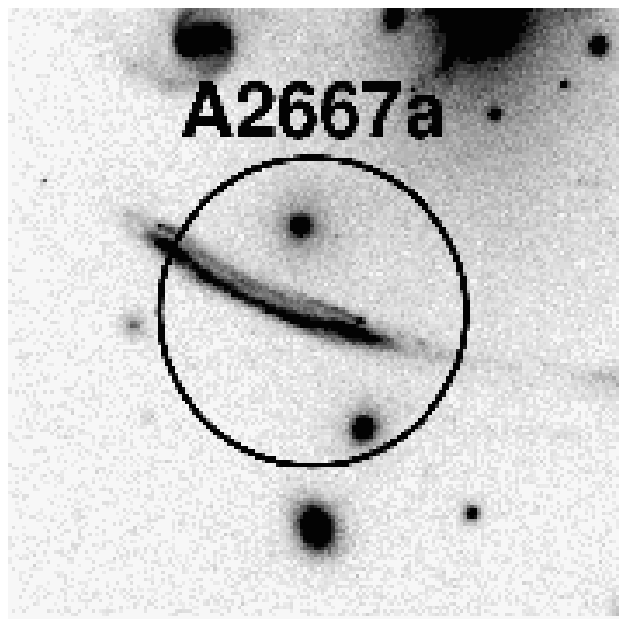} %{A2667a-s.ps}
\figcaption{Postage stamps of the lensed galaxies. 
Images are from the Hubble Space Telescope, 
Advanced Camera for Surveys,  filter F850LP,
with the following exceptions:
the MS0451a image is HST ACS F814W; 
the A2667a and A851a images are HST WFPC2 F814W;
and for A2261a and A2219a, the images are from
the Steward Observatory 90 inch, F606W.
Each postage stamp is 12\arcsec\ by 12\arcsec.
Circles of $R=6$\arcsec\ are drawn to 
illustrate the FWHM beam of Spitzer at 24~\micron.
}
\label{fig:postage}
\end{figure}

Since submillimeter sources in lensing clusters have been 
studied extensively, with published amplifications and redshifts, 
we prioritized  sources in the target list with published sub-mm detections.
Our sample contains 15 galaxies, targeted in GTO campaigns 82 and 30775.
Eight of these galaxies have sub-mm detections in the literature, as
referenced in Table~\ref{tab:sample}.
Figure~\ref{fig:postage} shows optical--band images of the sample
galaxies.

We supplement the sample with four  sources from the literature.
We use three of the four sources from \citet{teplitz07},
with  \fnu(24~\micron)$=$ 0.13--0.2~mJy; these are unlensed 
galaxies at $1<z<3$ with very long (9--12~hr)  IRS exposures, and
are not known sub-mm sources.
We also use the lensed $z=2.81$ sub-mm source in 
A370 from \citet{lutz05}\footnote{The two other sources 
from those papers,  ``2-x'' from \citet{teplitz07} and 
the Abell 2125 source from \citet{lutz05}, have insufficient 
signal-to-noise for our analysis.}  
with \fnu(24~\micron)$=$1.36~mJy.

Source coordinates, redshifts, amplifications, and flux densities
or our sample and for the four literature sources 
are all listed in Tables~\ref{tab:sample} and Table~\ref{tab:longwave}.

\section{Observations and Data Reduction}

This paper is based on a combination of imaging with MIPS, 
spectroscopy with IRS, archival imaging from Chandra, and 
submillimeter photometry from the literature. 
Exposure times are summarized in Table~\ref{tab:exptime}.

\subsection{Spitzer/IRS spectroscopy}
Low-resolution IRS spectra were obtained as part of 
Spitzer GTO programs 82  and 30775 (PI G.~Rieke).
Data reduction used a package
developed by D.~Fadda, which has already been successfully used
to reduce low-resolution IRS spectra of faint high-z sources 
\citep{yan07}.  Residual background, 
rogue pixels\footnote{http://ssc.spitzer.caltech.edu/irs/roguepixels/} 
(pixels with dark current
values abnormally high and variable with time) and cosmic ray hits
have to be corrected to obtain an optimal reduction of the data.  The
package considers all the bidimensional frames produced by the IRS/SSC
pipeline. First, background and noise images are produced by coadding
frames after masking target and serendipitous spectra on each frame. 
(The background is better estimated for the PID 30775 observations
where both LL1 and LL2 orders provide redundant background measurements,
compared to the PID 82 observations using only LL1, where the background
can only be determined by differencing nods and thus suffers from
contamination by  faint sources.)

A robust statistical estimator (biweight) is used to
minimize the effects of deviant pixels on the coadded value.  This part
is iterative in order to allow manual identification of sources.
Rogue pixels are then identified by computing the dispersion of the
noise around every pixel and flagging pixels which are
$> 5\sigma$ deviant from the mean local value.   Spectral extraction is done 
optimally (e.g., by using the PSF profile to weight the spectrum), 
taking into account the spectral distortion, and also rejecting
pixels affected by cosmic rays.  Finally, the spectra obtained at each 
position are co-added.
%[**Add how LL1 was used to estimate LL2 sky, and vice versa.]

%NAMES FOR THE 3 images:
% Kneib et al 2004      NED   
% A  SMM J16359+6612.6  J163555.2+661238
% B  SMM J16359+6612.4  J163554.2+661225
% C  SMM J16358+6612.1  J163550.9+661207

We downloaded and reduced one archival IRS spectrum, 
from program  3453 (PI P.~van der Werf).  
This is \imageA, image ``A'' in \citet{kneib04}, one of three
images of the same triply--imaged sub-mm galaxy in Abell~2218.
We obtained an IRS spectrum of the brightest image of this galaxy, 
\imageB\ (image ``B'' in \citealt{kneib04}); its 24~\micron\ flux
is 1.8 times brighter than the \imageA.
In the archival spectrum, the source was mis-centered across the short 
axis of the slit, which reduced throughput.  Because its signal-to-noise
ratio is low, we use the archival spectrum only 
to confirm the detection of features, and use the spectrum from our program
for all analysis.

\begin{figure}
\figurenum{2}
\includegraphics[width=6in,angle=0]{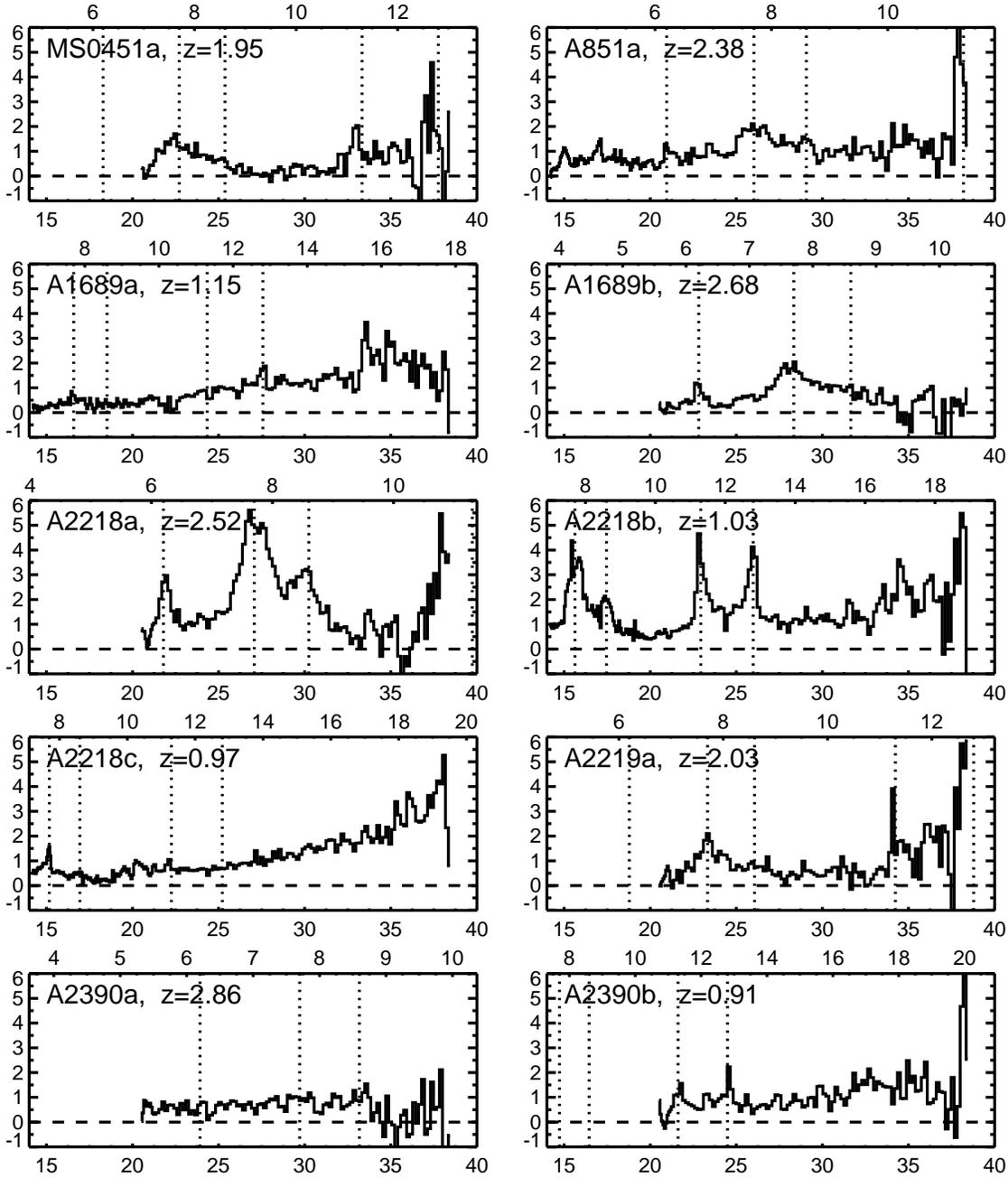}
%{/WORK/Clusters/IRS_spectra_lensed/Plots/Eiichi_spec/p1.ps}
%{/WORK/Clusters/IRS_spectra_lensed/Plots/Eiichi_spec/plot_batch2.epsi}
\figcaption{IRS spectra.  Lower x-axes show observed wavelength in \micron; 
upper x-axes show rest wavelength (\micron) if redshift is known.  
Y-axes plot observed flux density (mJy).
Spectra obtained for the same object in different programs 
are plotted separately, with program 82 plotted first.
Vertical dotted lines show the positions of 
expected aromatic and fine structure emission features.
Note that artifacts are present at $\lambda > 34$~\micron.
}
\label{fig:allspec}
\end{figure}

The IRS spectra are presented in Figure~\ref{fig:allspec}. 
Spectra obtained in programs 82 and 30775 for the same object are plotted
separately; observations from the latter program have superior background
subtraction.
Flux densities are as observed, without any correction for 
gravitational amplification.
When redshifts were known from the literature, or are evident from aromatic
emission features in the IRS spectrum, then the upper x-axis shows
rest wavelength.  Only one source (A2261a) lacks a redshift.
Redshifts are listed in Table~\ref{tab:sample},
with a reference if the redshift is from the literature.  

The rest--frame spectra were run through PAHFIT \citep{jdsmith07} 
to fit the continuum and aromatic (PAH) features simultaneously.
Measured aromatic fluxes are reported in Table~\ref{tab:pahfluxes}.

We allowed PAHFIT to fit the silicate absorption 
(e.g., we did not specify ``NO\_EXTINCTION'').  Fixing the
silicate optical depth at zero changes the best-fit 7.7~\micron\
feature fluxes by $4\%$ (median), the best-fit 8.6~\micron\
feature fluxes by $8\%$ (median), and the best-fit 11.3~\micron\
feature fluxes by $11\%$ (median).

We report fluxes in table~\ref{tab:pahfluxes} for features in which 
the measured flux is at least twice the uncertainty calculated for 
that line.  We add back in a few features, mostly at 11.3~\micron,
which are obviously detected even though the errorbars are large.

\subsection{Spitzer/MIPS 24~\micron\ and 70~\micron\ photometry}
MIPS photometry--mode images at 24~\micron\ and 70~\micron\ 
were obtained through Spitzer GTO program 83 (PI G.~Rieke).  
Exposure times  are given in Table~\ref{tab:exptime}
and photometry is given in Table~\ref{tab:longwave}.
The data were reduced and mosaicked using the 
Data Analysis Tool \citep{gordon06} with a few additional
processing steps \citep{egami-bcg06}.
% give pixel scale for 70 um!

Photometry at 24~\micron\ was obtained by PSF fitting, using the 
IRAF implementation of the DAOPHOT task \textit{allstar}.  
The PSF was created empirically using all available images from program 83.
An aperture correction of 1.131 was applied, calculated from a
Tiny Tim model of the 24~\micron\ Spitzer PSF that extends 
to $r=220$\arcsec\ (C.~Engelbracht, priv.~comm.)\footnote{
The aperture correction is specific to the DAOPHOT 
parameters used, in this case sky annulus radii of $31.3$\arcsec\ 
and $40$\arcsec, and a PSF defined out to $r=22.41$\arcsec.}

Aperture photometry at 70~\micron\ was performed on the 1.0 pixel scale
images.  The aperture radius was 16\arcsec\ 
(twice the half-width at half-max), 
the background annulus had radii of 18\arcsec\ and 39\arcsec\ 
(the position of the first airy ring), and an aperture correction 
of 1.968 was applied.\footnote{http://ssc.spitzer.caltech.edu/mips/apercorr/}
If the source was undetected, we took the upper limit as 
$2\times$ the $1 \sigma$ sky noise plus any positive flux in 
the source aperture.

\subsection{Spitzer/IRAC photometry}
IRAC images at 3.6, 4.5, 5.8, and 8~\micron\ were 
obtained as part of GTO program 83 (PI G.~Rieke) for five of the six clusters;
IRAC data for Abell 1835 were obtained through GTO program 64 (PI G.~Fazio).
Exposure times are listed in Table~\ref{tab:exptime}.
IRAC images were mosaicked as described by \citet{huang04}.

Photometry was obtained by aperture photometry with sky subtraction.
The high density of cluster galaxies in the cluster cores 
necessitated use of 
irregularly--shaped polygons for both target and sky apertures.  
The flux density within these polygonal regions was determined using 
CIAO\footnote{http://cxc.harvard.edu/ciao/} Versions 3.2.1 and 
3.4.1.1.
IRAC photometry is listed in Table~\ref{tab:longwave}; the values
do not include aperture corrections.
%We have not applied aperture corrections, but presumably should.

\subsection{HST photometry from WFPC2 and ACS}
From the HST archive, we downloaded all publically available ACS images for
the clusters.  We used Multidrizzle \citep{multidrizzle}  to 
distortion--correct each flat-fielded pipeline image, and then 
cross--correlated the images to measure the small 
($<1$\arcsec) coordinate registration offsets with high precision.
The measured offsets were then used by multidrizzle for the final mosaicking.  
The data were photometered using the same technique as for the IRAC bands.

We also used reduced WFPC2 images kindly made available by D.~Sand, 
which are described in \citet{sand05}.
ACS and WFPC2 cutouts of the galaxies in our sample are shown in 
Figure~\ref{fig:postage}.

\subsection{Chandra ACIS imagery}
\label{sect:xraydata}
From the Chandra archive, we downloaded all publically available 
ACIS observations of clusters with IRS targets in this sample:
observation ID numbers
\dataset [ADS/Sa.CXO#obs/00529]  {529}  and 
\dataset [ADS/Sa.CXO#obs/00902]  {902}  for MS0451;
\dataset [ADS/Sa.CXO#obs/01663] {1663}, 
\dataset [ADS/Sa.CXO#obs/05004] {5004}, and 
\dataset [ADS/Sa.CXO#obs/00540]  {540}  for A1689;
\dataset [ADS/Sa.CXO#obs/00495]  {495}  and 
\dataset [ADS/Sa.CXO#obs/00496]  {496}  for A1835;
\dataset [ADS/Sa.CXO#obs/01454] {1454},
\dataset [ADS/Sa.CXO#obs/01666] {1666}, and
\dataset [ADS/Sa.CXO#obs/00553] {553} for A2218;
\dataset [ADS/Sa.CXO#obs/00896] {896} for A2219;
\dataset [ADS/Sa.CXO#obs/05007] {5007} and
\dataset [ADS/Sa.CXO#obs/00550] {550} for A2261; 
\dataset [ADS/Sa.CXO#obs/04193] {4193},
\dataset [ADS/Sa.CXO#obs/00500] {500}, and
\dataset [ADS/Sa.CXO#obs/00501] {501}, for A2390; and
\dataset [ADS/Sa.CXO#obs/01562] {1562} for AC114.

%1562for   AC114.
%529, 902        for   MS0451;  
%1663, 5004, 540for   A1689;   
%495, 496for   A1835;   
%896for   A2219;   
%5007, 550for   A2261;   
%4193, 500, 501for   A2390;   and

We used CIAO (3.3) to 
update the CTI correction when necessary and remove data obtained during 
periods of high solar flares.
We mosaicked multiple overlapping observations following the 
CIAO thread ``Reprojecting Images: Making an Exposure--corrected Mosaic''.

For X-ray detections, fluxes were determined from the individual 
(not mosaicked) observations as described in \citet{delphine07}.
X-ray flux upper limits ($3 \sigma$) were determined from the 
mosaicked images and mosaicked exposure maps as in \citet{jennradio}, 
except that the $90\%$ encircled energy radius was used, 
and the Monte--Carlo sky apertures were taken close
to the source (within 30\arcsec) in order to sample
a representative cluster background.

Chandra X-ray fluxes and upper limits for the 2--8~keV and 0.5--8~keV
bands are reported in Table~\ref{tab:xray}.

\subsection{Long--wavelength photometry from the literature}
\label{sec:litphot}
Eight galaxies in our sample, and one galaxy from the extended 
(literature) sample, have submillimeter detections in the literature;
these are listed and referenced in Table~\ref{tab:sample}.
Four sources were detected at 15~\micron\ by ISO:
A851a and A2218b \citep{barvainis}, A2390b \citet{lemonon98}, and 
A2390c \citep{metcalfe}. 
We will use this photometry to help determine  the total 
infrared luminosities of the sources.

\section{Discussion}
\subsection{Aromatic Feature Flux Ratios in High--Redshift Galaxies}

We now test whether the mid-IR spectra  of star--forming
galaxies, with their prominent aromatic emission features, show 
evidence for evolution from $z=0$ to high redshift.  

\begin{figure}
\figurenum{3}
\includegraphics[width=5in,angle=270]{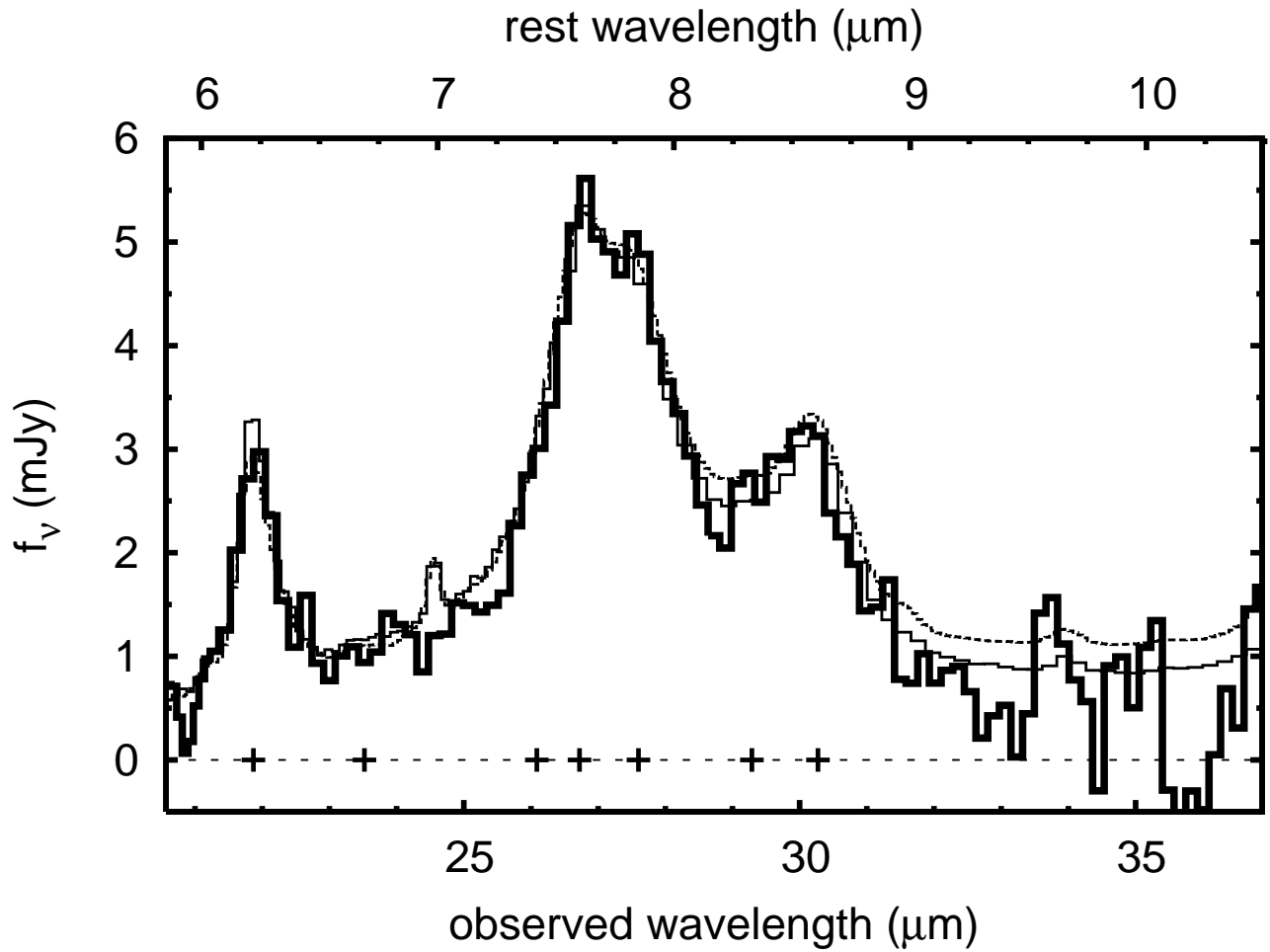}
%{/WORK/Clusters/IRS_spectra_lensed/Plots/A2218/Early_plots/combine.ps}

\figcaption{IRS spectrum of $z=2.516$ lensed source \imageB\ 
behind Abell 2218  \textit{(solid thick line)}.
Overplotted are two $z \sim 0$ spectral templates:
the  starburst galaxy NGC~2798 from \citet{dale06} \textit{(thin solid line)};
and the average of 13 starburst galaxies from 
\citet{brandl06} \textit{(thin dashed line)}.
Crosses show the wavelengths of known aromatic components from \citet{jdsmith07}.
}
\label{fig:a2218_spec}
\end{figure}

Given its high quality, we first examine the IRS spectrum for 
\imageB\ in detail.
The image is amplified by a factor of $22 \pm 2$
\citep{kneib04}, such that
without lensing, its 24~\micron\ flux density would have
been 53~\microJy, at the MIPS confusion limit.
The galaxy redshift, from H$\alpha$, is $z=2.5165 \pm 0.0015$ \citep{kneib04}.
Its IRS spectrum, presented in Figure~\ref{fig:a2218_spec}, 
shows highly--significant detections of aromatic features at 
6.22, 7.7, 8.33, and 8.61~\micron.
(The lower-quality spectrum from program 3453 confirms the 
detection of these aromatic features, though the two longer--wavelength 
features are detected at low significance.)
Overplotted with arbitrary normalization 
are two spectral templates:  a) the starburst galaxy NGC~2798
from the SINGS survey \citep{dale06};
and b) the average spectrum of 13 nearby starburst galaxies 
with little apparent AGN contribution \citep{brandl06}, 
whose mean IR luminosity is $4.9 \times 10^{10}$~\Lsun.
Coincidentally, this is also the L(TIR) of NGC~2798.

The $z=2.516$ spectrum  in Figure~\ref{fig:a2218_spec} 
closely resembles the low--redshift, lower--luminosity 
(by a factor of 15) templates.
Aromatic flux ratios are reported in Table~\ref{tab:pahrats}, 
along with those we measure for the Brandl template, 
and those reported for the SINGS sample.
The strength of 
the 8.6~\micron\ feature in \imageB\ relative to those at 
6.2~\micron\ and 7.7~\micron,  
is  within the $10\%$--$90\%$ variation within the SINGS sample. 
In addition, the flux ratios of \imageB\ are a close match to those of the 
\citet{brandl06} average starburst template;
while 7.7~\micron\ is relatively stronger than the template, 
the difference is comparable to the standard deviation of 
line ratios for individual galaxies that were averaged 
to create the template.
{\bf Thus, we conclude that this $z=2.5$ galaxy has aromatic feature flux ratios 
that are  consistent with those observed for lower--luminosity, 
$z=0$ starbursting galaxies.}

\begin{figure}
\figurenum{4}
\epsscale{0.9}
\plottwo{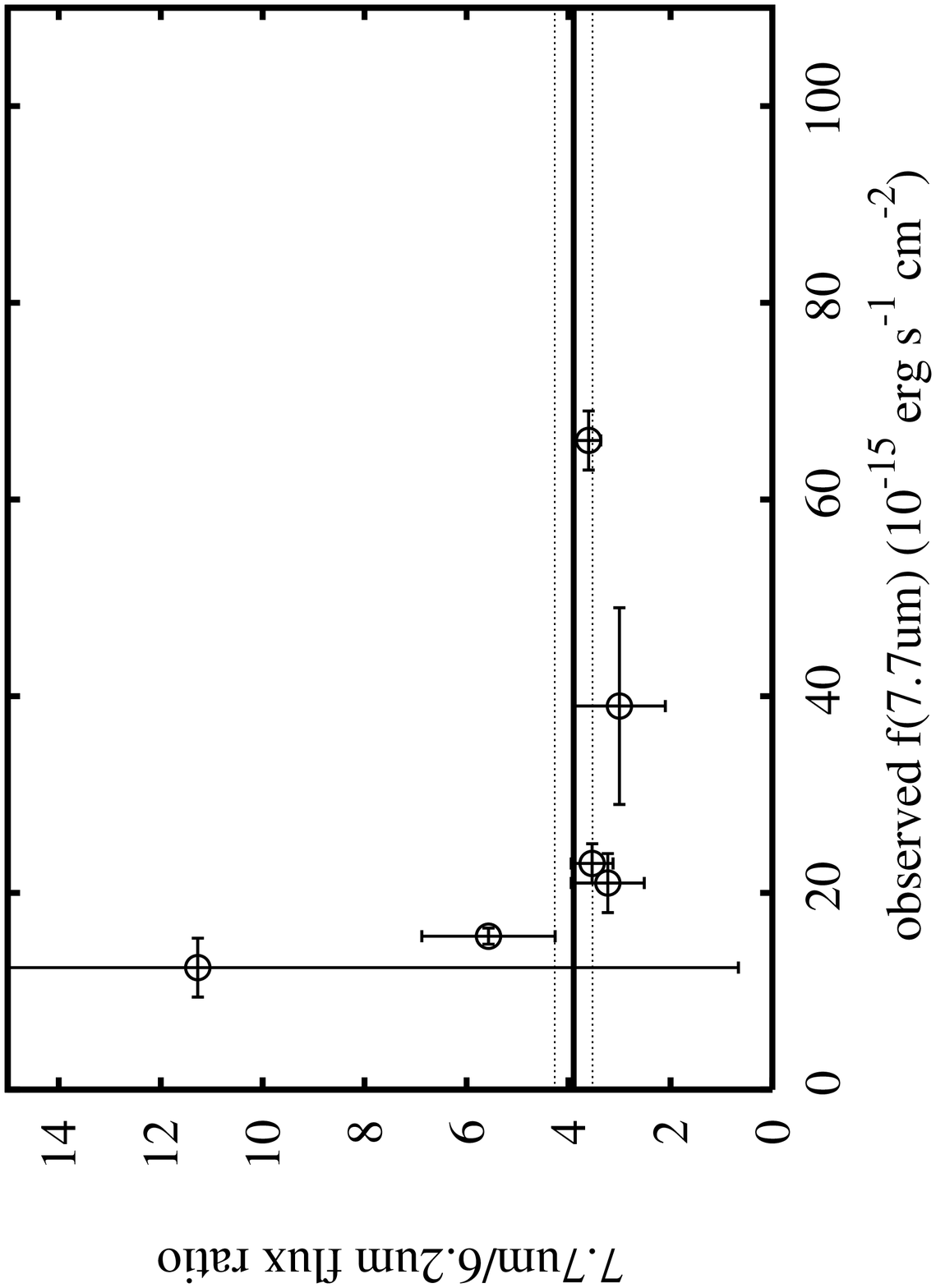}{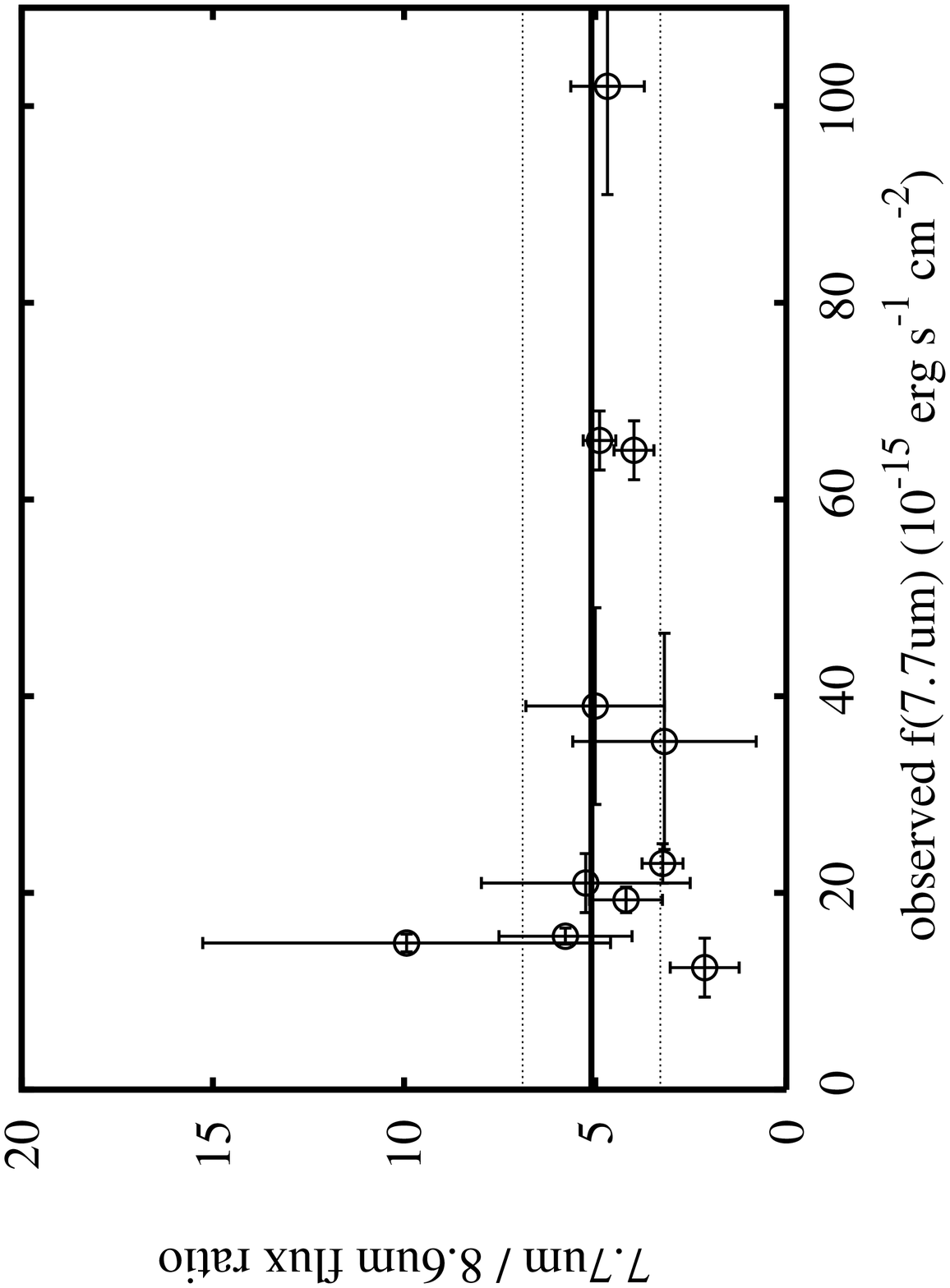}
%{/WORK/Clusters/IRS_spectra_lensed/Other_IRS_reductions/Eiichi-Fadda/Pahfit/rat1.ps}
%{/WORK/Clusters/IRS_spectra_lensed/Other_IRS_reductions/Eiichi-Fadda/Pahfit/rat3.ps}

\plottwo{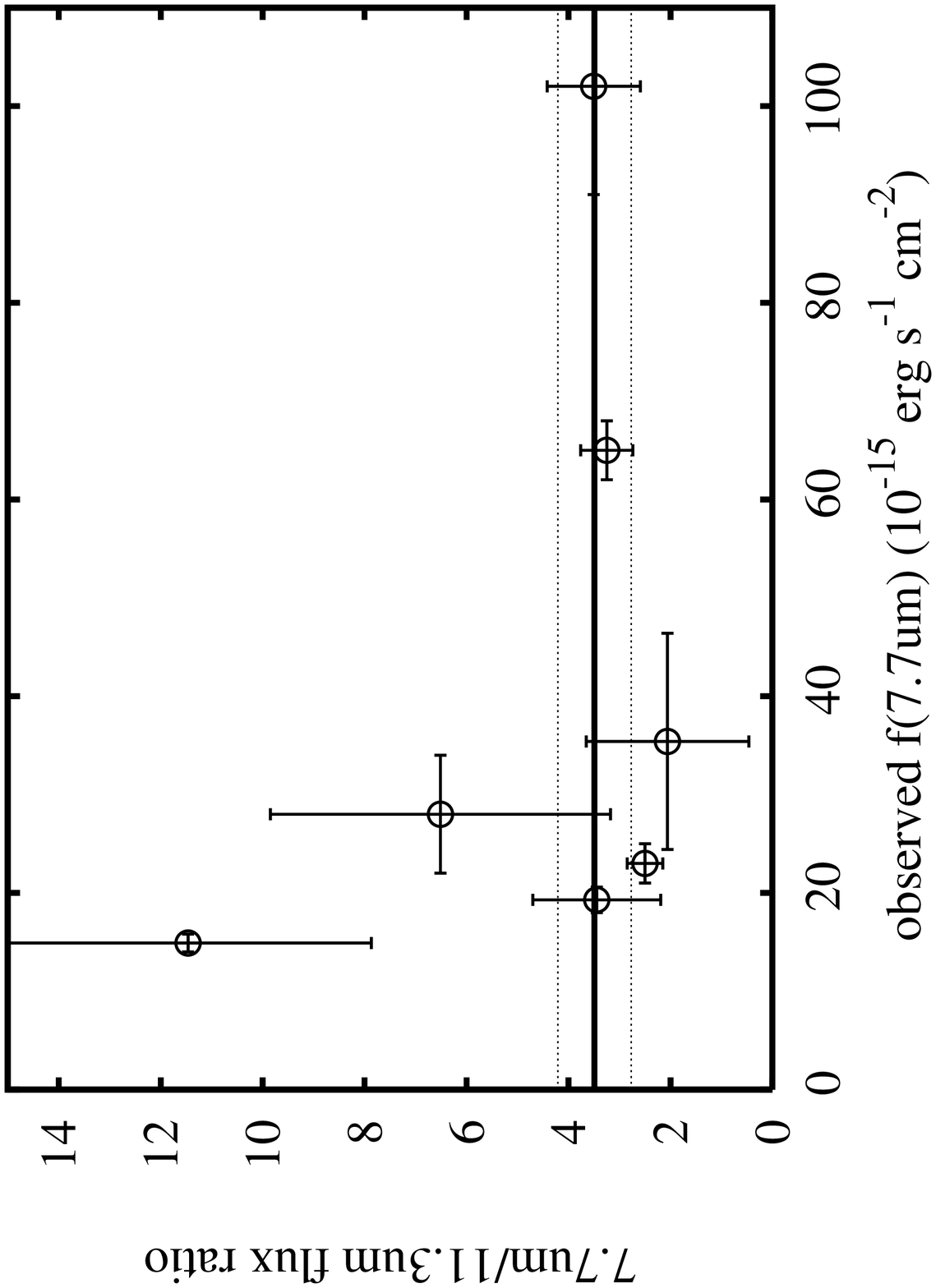}{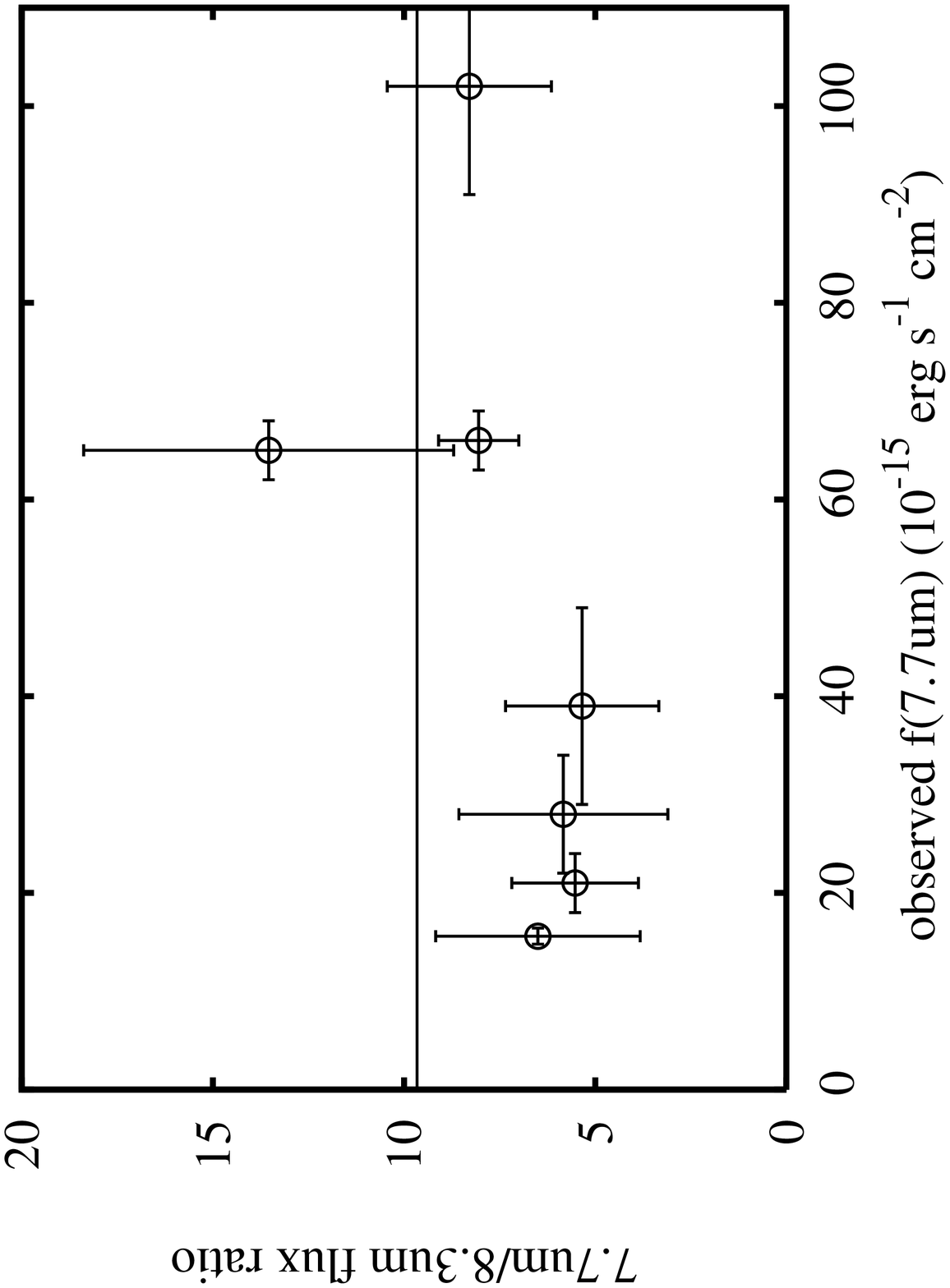}
%{/WORK/Clusters/IRS_spectra_lensed/Other_IRS_reductions/Eiichi-Fadda/Pahfit/rat2.ps}
%{/WORK/Clusters/IRS_spectra_lensed/Other_IRS_reductions/Eiichi-Fadda/Pahfit/rat4.ps}
\epsscale{1.0}
\figcaption{Aromatic flux ratios.    The aromatic flux ratios are plotted against 
observed flux in the 7.7~\micron\ feature.    Fluxes were fit using PAHFIT.
Amplifications have  not been divided out.   Also plotted are the 
flux ratios of the low--redshift \citet{brandl06} starburst template
\textit{(solid horizontal lines)}, with the $\pm 1$ standard deviations from that
sample \textit{(dotted horizontal lines)}.
}
\label{fig:pahrats}
\end{figure}

While the other spectra in the sample have lower signal-to-noise 
ratios, they are still sufficient, in the aggregate, to test
whether high--redshift galaxies have markedly different aromatic flux
ratios.  Figure~\ref{fig:pahrats} compares, for four aromatic feature flux ratios,
the observed ratios in our sample with those measured
for the low--redshift \citet{brandl06} template.  
Though the errorbars in individual measurements are sizable,
Figure~\ref{fig:pahrats} confirms the result seen for \imageB:
star--forming galaxies at $1<z<3$ have aromatic feature
ratios consistent with those of $z=0$ star--forming galaxies.  
%There is no evidence for strong evolution in aromatic feature flux ratios.

Variations in aromatic feature ratios have been reported in certain
HII regions---this may be attributable
to variations in size, composition, and ionization state
of the carriers  (\citealt{draineli07} and references therein).
In particular, large PAHs emit more strongly at 11.3~\micron, 
and neutral PAHs emit
much more strongly at 3.3 and 11.3~\micron\ than charged PAHs
\citep{draineli07,galliano06,allam89}.  
That our measured feature ratios at $z>1$
are consistent with those measured at $z\sim0$ suggests 
that the dust grain size and ionization distributions are not 
strongly evolving.

\subsection{Evidence for Compact Source Accretion}
\label{sec:nuc}

Active galactic nuclei are detectable in multiple ways---by
the dust they heat that radiates in the mid-IR;
by X-rays emitted from the accretion disk coronae;
and by high--excitation or broad emission lines.
We now examine the evidence for AGN activity in our extended sample.

\begin{figure}
\figurenum{5}
\includegraphics[width=4in,angle=270]{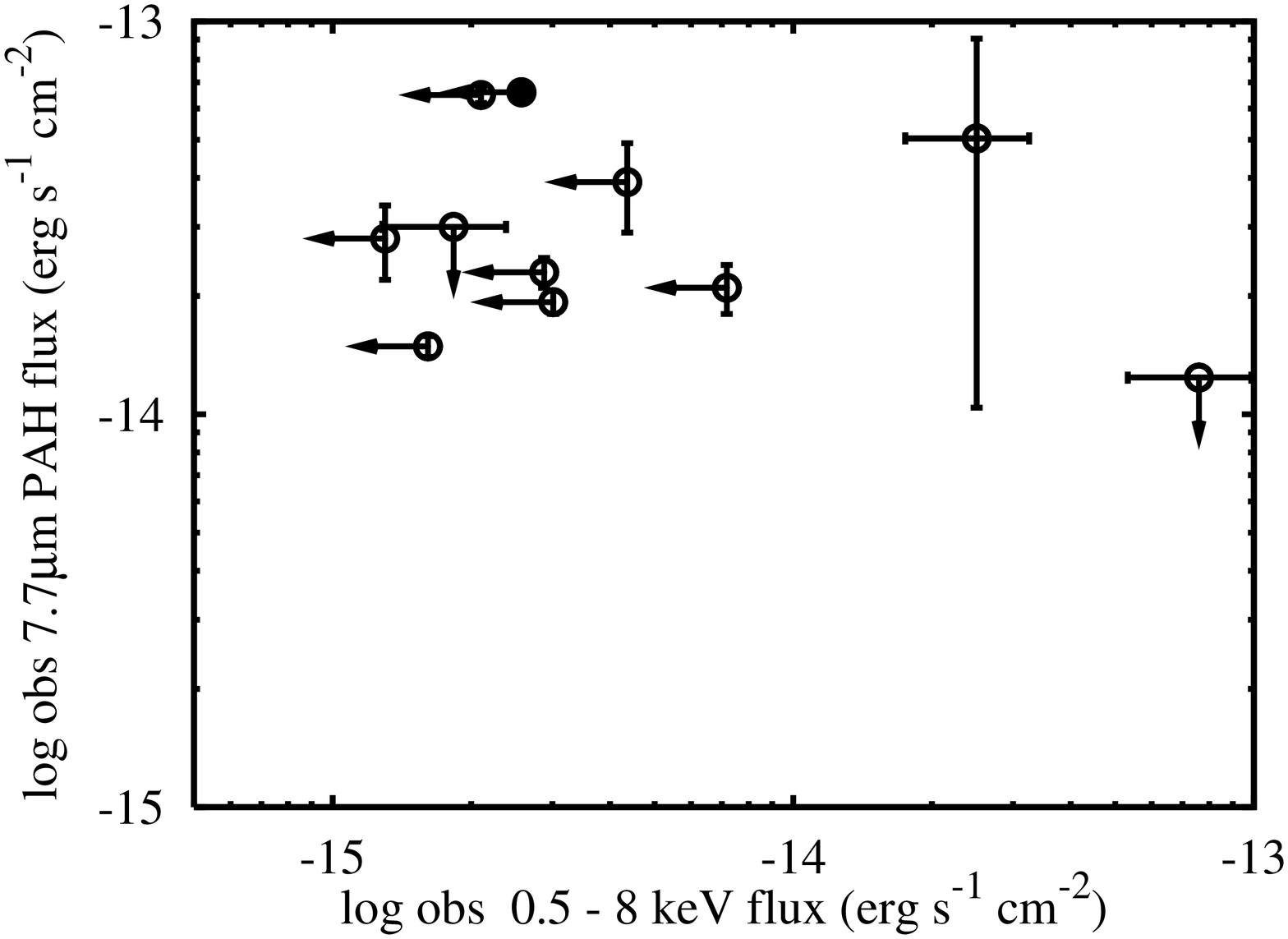}
%{/WORK/Clusters/IRS_spectra_lensed/Xray_limits/Plots/new-f7.7xray.ps}
\includegraphics[width=3.9in,angle=270]{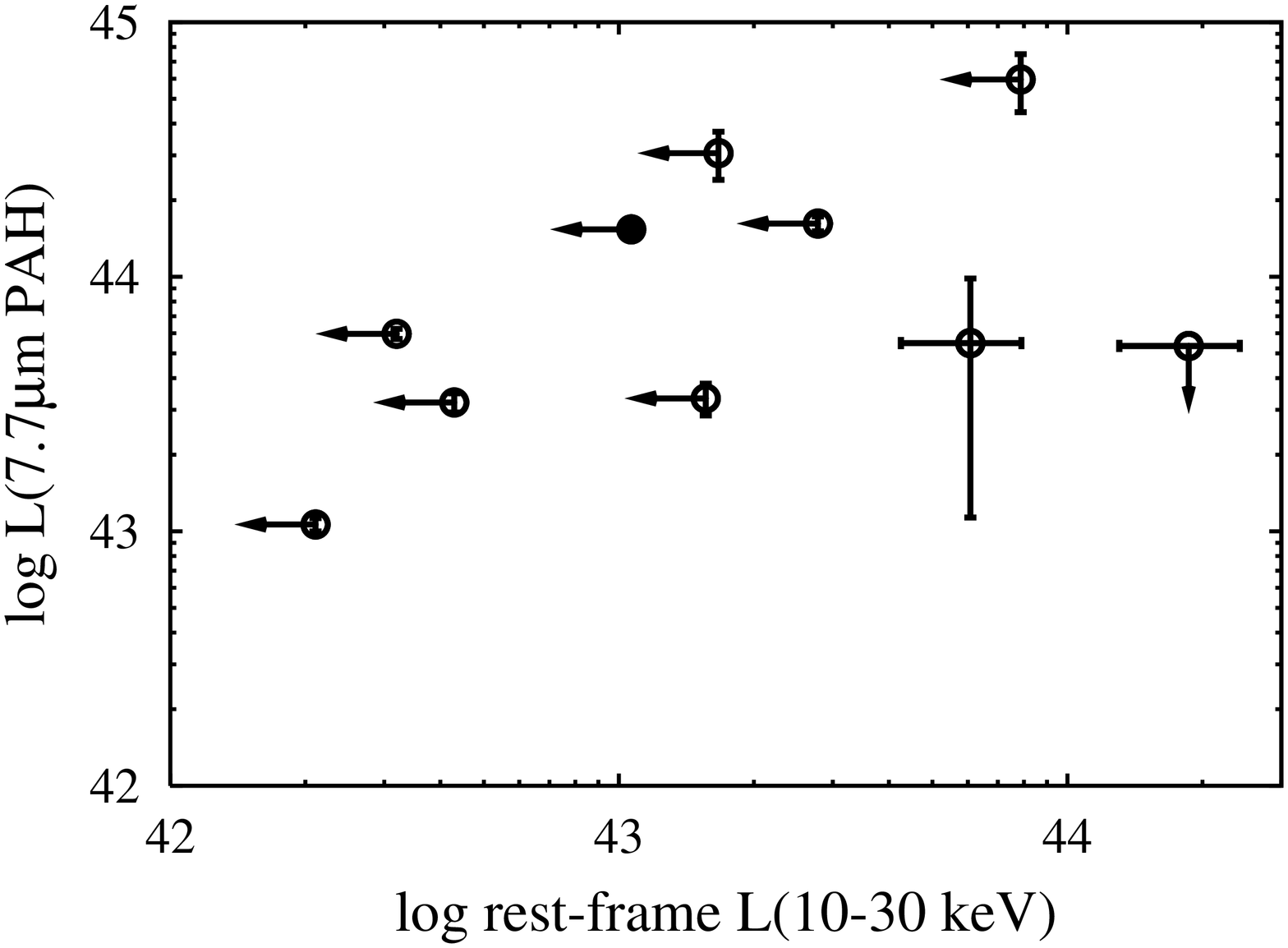}
%{/WORK/Clusters/IRS_spectra_lensed/Xray_limits/Plots/new-L7.7-xray.ps}
\figcaption{Comparison of aromatic and X-ray fluxes and luminosities.
\imageB\ is represented by filled symbols.
}
\label{fig:fpahX}
\end{figure}

The first AGN diagnostic we examine is X-ray luminosity, plotted 
in Figure~\ref{fig:fpahX} against aromatic luminosity.
Most of the sample are X-ray non--detections, at 
limiting luminosities that rule out X-ray--loud QSOs or bright Seyferts,
although an X-ray--weak or highly obscured AGN could still be present
(c.f. \citealt{jennradio,almu-plagn}).
%\imageB \textit{(filled circle)}  certainly has more luminosity
%emitted in the 7.7~\micron\ aromatic feature alone than is 
%expected for an AGN from the X-ray upper limit.
Two of the four galaxies from the literature, and three of the
galaxies in our program are detected in X-rays:
two have no apparent aromatic features (sources A2261a 
and A2390a); and the other, source A2390b,  
is interesting in that it contains a 
luminous X-ray--emitting AGN, yet its spectrum still shows aromatic features.
That source emits roughly equal power in the 7.7~\micron\ aromatic feature
(scaled from the 11~\micron\ feature) and at 10--30~keV,
reminiscent of the more luminous $z=1.15$ source 
CXO GWS J141741.9+522823 discussed by \citet{lefloch07}.

We now consider a second AGN diagnostic, the relative contribution 
of aromatic versus continuum emission to the mid-IR flux.  
At low redshift, low aromatic feature equivalent widths  have been demonstrated
as an effective AGN diagnostic (\citealt{armus07,brandl06}, and
Devost \etal\ (in prep.)).  
However, it is extremely difficult to measure 
equivalent widths accurately 
for the high--redshift IRS sources, due to the limited wavelength
baseline,  the fact that the aromatic features have broad wings,
and most importantly, imperfect sky subtraction which adds a
(positive or negative) pedestal to each spectrum.
For example, for \imageB\ (our highest--quality spectrum), 
we cannot measure an accurate equivalent width due to the 
difficulty in determining the true continuum level.  
We are able to set a lower limit 
of  $>0.4$~\micron\ (rest-frame, conservatively assuming a high continuum level),
which is a typical value for low--redshift star--forming galaxies.  
For the other spectra in the sample, equivalent widths are even harder
to measure.
%pahfit gives EW(6.2um)r = 2.5.  that's very high.  
% Because it's fit zero continuum.

Therefore, we instead create an aromatic-to-mid-IR flux ratio,
which we define as the ratio of the flux in the 7.7~\micron\ PAH feature
(in \cgsflux, as fit by PAHFIT) to the MIPS photometry at 24~\micron\ 
(in Jy, corrected for bandwidth compression to $z=0$.)
This metric does not suffer as strongly from the difficulty in
determining the continuum level.  In the low--redshift, star--forming 
comparison sample we construct in \S~\ref{sec:pahtotrat}, this ratio ranges from
7--32 $\times 10^{-11}$.
We thus take $7 \times 10^{-11}$ as a dividing line between 
spectra dominated by star formation (above the value), and spectra
with substantial AGN contribution (below the value).
This discriminant selects the following mid-IR spectra\footnote{For A2390b, 
we scale from the flux in the 11~\micron\ aromatic feature,
using the average 7.7/11~\micron\ flux ratio of 3.6 from \citet{jdsmith07}.}
as having substantial AGN contribution:
A370a, MS0451a, A1689a, A2218c, A2261a, A2390a.
% A2219a just squeaks by
Thus, of 19 galaxies with adequate spectra, 6 show strong indications
of AGN contribution to their mid-IR outputs.

A third indication of AGN activity is a mid-IR spectrum that rises steeply with 
increasing wavelength.  Four spectra (A2261a, A2218c, A1689a, and A370a) show
this behavior; none has detected aromatic emission.

A fourth indication of AGN activity is the presence of broad or
highly excited lines in an optical spectrum.
Of the spectroscopy that has been published, or is in preparation, for our sample
(see references in Table~\ref{tab:sample}), 
three show evidence for AGN:
A1689a has a Keck spectrum showing highly ionized neon 
(J.~Richard \etal\ in prep.);
A2390a has a Keck spectrum (see Appendix) which shows Lyman $\alpha$ with FWHM 
$\sim 945$~km~s$^{-1}$, typical of an AGN narrow line region; 
and the spectrum of A370a is reported to contain AGN lines \citep{lutz05}.

A fifth indication of AGN activity is a high ratio of 
[Ne III] 15.5~\micron to [Ne II] 12.8~\micron.  Solar--metallicity
starbursts are not observed to excite this line ratio above unity
\citep{thornley,rigby-rieke}, whereas the harder continuum of AGN do.
Eight spectra cover rest-frame 12.8\micron; [Ne II] is detected in 6 of these.
Six spectra cover rest-frame 15.5~\micron;  [Ne III] is detected in 4.
Of the four galaxies where both lines are detected 
(A1689a, Ab2218b, A2390b, and A2667a), only in A1689a does the [NeIII]/[NeII]
ratio exceed unity:  $2.15 \pm 0.76$.   This high line ratio offers
additional evidence that A1689a hosts a luminous AGN.
The other, lower line ratios are more difficult to interpret.

The results for these AGN diagnostics 
are compiled in Table~\ref{tab:agn-diag}.  
Combining these diagnostics,
8 galaxies out of the sample of 19 have at least one
indication of AGN activity (X-ray detection, low aromatic 
contribution to the mid-IR flux, rising mid-IR spectrum, or 
optical AGN lines).  Five galaxies  have two different diagnostics that
indicate AGN.  
Thus, we find that $42\%$ of the faint 24~\micron\ sources in this sample 
have some nuclear activity; the $95\%$ confidence interval is $23\%$--$63\%$.
Further, in $32\%$ of the sample, the AGN  strongly affects or dominates 
the mid-IR spectrum; the $95\%$ confidence interval is $15\%$--$53\%$.
(Confidence intervals were calculated using eqn.~26 of \citet{gehrels86}.)
% calculated in  gehrels_poisson_binomial_errorbars.ods

The high proportion of AGN in our sample of high--redshift LIRGS and ULIRGs
mirrors the general behavior of local IR-luminous galaxies, where AGN
become substantially more common with increasing luminosity
(e.g., \citealt{lutz98}).
Though the errorbars are large given the small sample size, the
high proportion ($32\%$) of AGN sufficiently powerful to dominate the mid--IR
spectra may be inconsistent with the frequent assumption that 
$\sim 100$~\microJy, $z>1$ sources are dominated by star formation,
with AGN a modest contaminant.  Our result,
if confirmed with better statistics, has implications for
24~\micron--derived star formation rates in deep surveys, 
where it is often assumed
that the mid-IR outputs of high--redshift, faint galaxies are 
dominated by strong aromatic features from star formation.
It also implies that a higher fraction of the integrated 24~\micron\
flux on the sky may arise from accretion than previously thought.

\subsection{Total infrared luminosity}

We now determine the total infrared (TIR) luminosities of the lensed
sources, so that we may examine the dependence of aromatic feature emission
on total infrared luminosity, and how this behavior may evolve with redshift.

\subsubsection{\imageB}

\begin{figure}
\figurenum{6}
\includegraphics[width=4.5in,angle=270]{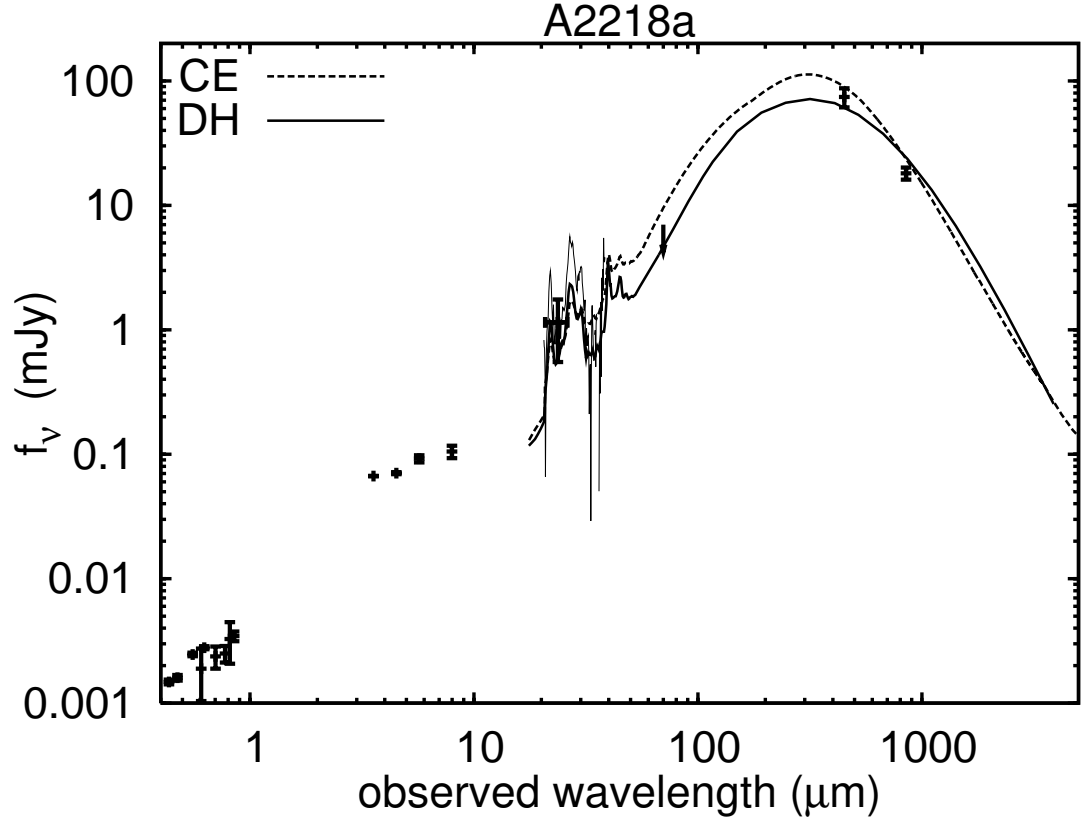}
%{/WORK/Clusters/IRS_spectra_lensed/Plots/A2218/Delphine/try.ps}
\figcaption{Photometry and L(TIR) fits.
Photometry and IRS spectra \textit{(thin lines)} are plotted, with 
flux densities and wavelengths in the observed frame.
Overplotted are best-fit templates: 
\citet{dalehelou}  (\textit{thick solid line}),  
\citet{charyelbaz}   (\textit{thick dashed line}), and Mrk 231 \textit{(dotted line)}.
IRAC and HST photometry are plotted for \imageB to demonstrate that its SED
is dominated by star formation.}
\label{fig:a2218_phot}
\end{figure}

The sub-mm galaxy behind Abell 2218 has exceptionally good 
photometric coverage from 0.4 to 850~\micron.  Since it is 
triply--lensed, we increase the signal-to-noise ratio of its
multiband  photometry by using fluxes from all three images, 
weighed by the amplifications given in \citet{kneib04}.
(Bright neighboring galaxies forced us to use only image A and B for
the optical and IRAC photometry.)
Figure~\ref{fig:a2218_phot} plots this photometry.

We fit redshifted \citet{charyelbaz} and \citet{dalehelou}
templates to the 24, 70, 450, and 850~\micron\ photometry, 
corrected for lensing amplification.  For each template family,
we determine the best--fit template and the range of acceptable 
templates, then determine L(TIR) for those templates in a method
consistent with the definition given in Table~1 of 
\citet{sanders-mirabel} using 12, 25, 60, and 100~\micron\ photometry.
We estimate the intrinsic total infrared luminosity of \imageB\ as:
\begin{itemize}
\item  $7.6 \pm1.9 \times 10^{11}$~\Lsun\ for \citet{dalehelou}  templates.
\item  $1.3^{+1.0}_{-0.2} \times 10^{12}$~\Lsun\ for \citet{charyelbaz} templates.
\end{itemize}

As Figure~\ref{fig:a2218_phot} shows,
the best--fit \citet{charyelbaz}  template over-predicts all $>24$~\micron\
points, especially at 70~\micron.   
The best--fit \citet{dalehelou} template
is a better match to the shape of the overall infrared SED
and also has high aromatic feature equivalent widths, consistent
with the spectrum of the source.
Thus, this galaxy's intrinsic luminosity is that of a LIRG.

\subsubsection{Other Galaxies}

We adopt a similar approach for the other galaxies with sufficient
far--infrared and submillimeter data to provide reasonable constraints.
We tried both \citet{dalehelou} and \citet{charyelbaz} templates, 
fitted to fall just below any upper limits so long
as they were compatible with the other constraints, i.e., to give a 
maximal luminosity.
Two galaxies are fit poorly by a star--forming template, and much better
by an AGN template, Mrk 231; MS0451a may be, as well.
%a star--forming template; we quote a conservative upper limit on L(TIR).
The ranges of acceptable luminosities and the preferred
templates are listed in Table~\ref{tab:longwave}. 
\citet{dalehelou} templates provided generally better fits to our sample
than \citet{charyelbaz} templates, as expected given the results 
of \citet{delphine06} for $0<z<1$ galaxies, fitting ISO, Spitzer, and 
VLA photometry.

%%%%%%%%%%%%%%%%%%%%%%%%%%%%%%%%%%%%%%%%%%%%%%%%%%
\subsection{8~\micron\ (rest)-to-Total Luminosity Ratio}  %George's new discussion
\label{sec:pahtotrat}

We now consider the aromatic and total infrared (TIR) luminosities of
the star--forming galaxies in the lensed sample, and how this ratio
behaves as a function L(TIR).  We compare to
higher--luminosity star--forming galaxies at $z\sim 2$, as well 
as comparable--luminosity star--forming galaxies at $z\sim 0.8$ and $z \sim 0$.
We interpret the properties of the high redshift
galaxies strictly in terms of properties that are well--determined for
local analogs, so we can test whether inconsistencies emerge.

\subsubsection{The $z \sim 0$ comparison sample}

To probe rest-frame 8~\micron\ and total infrared (TIR) luminosities
for local galaxies, we use the SINGS sample  
(photometry from \citealt{dale05})
supplemented in the ULIRG range by IRAS 09111-1007, 10565+2448, 
12112+0305, 14348-1447, 17208-0014, 22491-1808, and Arp220
%09111, 10565, 1211, 1434, 17208, 2249, and Arp 220 
\citep{farrah03,armus07},
all selected because they appear to be dominated by 
young stars rather than AGN.\footnote{IRAS 12112+0305 and 
14348-447 are double nuclei sources.}
We also include IRAS 00262+4251, 01388-4618, 02364-4751,
16474+3430, 23128-5919, and 23365+3604 from \citet{rigo99}. 
%Rigopoulou et al. (1999). 
The strength of the aromatic features relative to the continuum 
indicates that all these galaxies are also dominated by star formation.
For the galaxies from \citet{farrah03} and \citet{armus07}, we 
determined 8~\micron\ photometry from archival IRAC images.
In our reductions, we included a correction for the extended
source calibration.  For those galaxies in \citet{rigo99}, we converted 
the tabulated continuum and aromatic 7.7~\micron\
peak fluxes into 8~\micron\ photometric values by comparing with
templates for galaxies in common between \citet{armus07}:
%Armus et al. (2007): 
IRAS 12112+0305, 14348-1447, 1525+3609, and 22491-1808, plus Arp 220. 
The peak--to--peak scatter was a factor of two (in agreement with the
finding of \citet{charyelbaz}).
We add a variety of LIRGS \citep{almlirg},
again avoiding any galaxies with indications of
significant levels of AGN power. 
We took IRAS measurements from \citet{sanders03}, 
and computed L(TIR) to be consistent with the approach
taken in that paper. We define L(8~\micron) as being proportional to \nufnu, 
and use similar definitions for all luminosities at specific wavelengths.  
Figure~\ref{fig:TIRL8} shows the relation 
between rest--frame L(8~\micron) and L(TIR) for these $z\sim0$ galaxies.
%We fit it with the functional form proposed by 
%\citet{charyelbaz} for L(6.7~\micron) vs.\ L(TIR), normalized in 
%L(8~\micron) to fit the data for TIR luminosities below 
%$2 \times 10^{10}$ \Lsun.

\subsubsection{Results from the literature at $z\sim$ 0.85}
We add a sample of infrared galaxies at 
$z \sim 0.85$ from \citet{delphine06}.
This paper utilizes 15~\micron\ data from ISO as well as 24~\micron\ 
data from Spitzer and, for a subset, radio data from the VLA; 
the authors found that the galaxies had rest--frame 
L(8~\micron)/L(12~\micron) ratios at the high end of the local galaxy dispersion.
We make the assumption that  L(12~\micron) correlates with L(TIR) 
in that sample, as it is observed to do well at $z=0$.  
We computed rest 8~\micron\ luminosities for the
members of the sample with 0.55 $<$ z $<$ 1.2 and adequate 15~\micron\ photometry,
and corrected them to the IRAC 8~\micron\ band by comparison with the IRS spectrum
of IRAS 2249, a local star-formation--dominated ULIRG.
We took L(TIR) from the \citet{dalehelou} template fits \citep{delphine06}. 
The results are shown in green in Figure~\ref{fig:newTIR}. 
%Under our assumption that rest--frame L(12~\micron) correlates with L(TIR), 
%the galaxies in the \citet{delphine06} sample are shifted away from the
%$z=0$ L(8~\micron)/L(TIR) trend line. 
%in the same direction  as we find for the $z=2$ galaxies.
%%

There is no mid-IR spectroscopy for these galaxies, but if they had strong 
mid-IR contributions from AGN we would expect substantial dispersion in the 
L(8\micron) to L(TIR) ratio.  Such a large dispersion is not seen.
%Since a large dispersion is not seen, we 
%believe this sample gives another indication of a modification of the IR SEDs at
%high redshift compared with local galaxies of similar luminosity.

\subsubsection{Results from the Literature at $z \sim 2$}

We now compare the behavior of the 8$\mu$m luminosity vs. L(TIR) at
$z \sim 2$ with that seen in local luminous galaxies.  We first discuss
three results from the literature, and then show how our new
measurements complement them.

{\bf First, we took the stacked $z \approx 1.9$ galaxy template of \citet{daddi05}}
and compared the ratio of observed 850~\micron\ to 24~\micron\ flux densities,
referred to the appropriate rest wavelengths for each of the local
star--forming galaxies where there is sufficient data to construct an accurate SED
template (IRAS 1211, 1434, 17208, 2249, and Arp 220).  Excluding Arp
220, for which the ratio is 54, the average value for the other four
galaxies is 14, whereas the \citet{daddi05} template indicates a value of
6.5.  Thus, the template indicates stronger emission at rest 8~\micron\
relative to the submm than is typical of local star--forming ULIRGs.

We also compare with total infrared luminosity derived
from submm measurements.  Since the rest wavelength for the observed
850~\micron\ point in the Daddi \etal\  template is 293~\micron, 
we construct the average ratio of L(TIR)/L(293~\micron) 
for local LIRGs and ULIRGs with 
$4 \times 10^{11} <  L(TIR) < 2.5 \times 10^{12}$~\Lsun\ 
and high quality measurements at 350--450~\micron\ 
(IRAS 09111, 1211, 1434, and 17208, Arp 220, and NGC 1614).
We correct the flux densities for the differing effective filter
bandwidths using the IRAS 2259 template.
We correct to 293~\micron\ assuming a spectral slope of -3.3,
determined from a larger sample of local ULIRG templates.  We find

L(TIR) $=$ ($98 \pm 40$)  L(293~\micron)

\noindent and then plot the Daddi \etal\ point as a red square 
 in Figure~\ref{fig:TIRL8}.
We did not fit the observations of lower luminosity
galaxies, because the scatter at lower
luminosities is very large, probably due to varying amounts of cold
and warm dust in the infrared-emitting objects.

{\bf Second, we make use of the stacking analysis carried out by \citet{papovich07}.}   
They report observed--frame 24~\micron\ and 70~\micron\ stacked 
photometry for galaxies that are both undetected in X-rays and 
have non--power--law IRAC SEDS, and are therefore assumed 
to be powered predominantly by star formation. 
We computed the rest 8~\micron\ and 24~\micron\ flux
densities for this category (since they quote values for $z = 2$, the
rest wavelengths are represented directly by the measurements).
We corrected the 8~\micron\
flux density (derived from that observed at 24~\micron) to the equivalent for
the IRAC band, again using the SED of IRAS 2249. 
Finally, we compute L(TIR) from the relation derived from our local galaxy sample:

%L(TIR) $=$ $86.53$ $\times $ L(24~\micron) $\wedge$ $0.975773954$  % eqn format
L(TIR) $=$ $86.5$ $\times $ L(24~\micron, rest)$^\alpha$, where $\alpha = 0.9758$.  % eqn format

The rms scatter around this relation is only $0.14$ dex.  The resulting
values are shown in Figure~\ref{fig:TIRL8} as red diamonds 
(we omitted the lowest flux density bin from Papovich \etal\ 
because the signal to noise ratio at 70~\micron\ is too low).

{\bf Third, we use the results of \citet{yan07}.}
We select those galaxies  that have $z>1.5$, 
and rest--frame 6.2~\micron\ equivalent widths (measured by \citealt{sajina07})
greater than 0.3~\micron.  
From \citet{armus07}, this equivalent width threshold should
select galaxies dominated by star formation.  The galaxies selected 
(MIPS 289, 8493, 15928, 16144, 22530) are all categorized as ``type 1'' 
by \citet{yan07}, meaning their spectra contain strong aromatic features.
The median rest-frame EW(7.7~\micron) is  2.5~\micron, corroborating that
star formation dominates the mid-IR spectra.
We converted the tabulated L(5.8~\micron) for these galaxies into L(8~\micron) 
from an average template for our local ULIRG sample (IRAS 2249).

We estimate L(TIR) for these sources using the radio-infrared
relation. We describe our adjustment of this relation for use at high
redshift in detail in a forthcoming publication (N. Seymour \etal, in
preparation) but give a short summary here. \citet{yun01} have
quantified the relation for a large number of local galaxies, finding
that the average ratio of flux density at 60$\mu$m to 1.4GHz is
110 independent of luminosity.  The results of \citet{yun01} have
not been K-corrected; since the infrared SEDs drop rapidly toward the
blue, while the radio spectrum drops slowly, these corrections are a
fairly steep function of redshift.  They enter at a significant level only for
the high luminosity galaxies, because they tend to be at higher
redshift than the lower luminosity ones. Including them increases the
indicated ratio to $\sim 170$ at a L(TIR) $= 10^{12}$ \Lsun. A
second issue is that the radio outputs of LIRGs and ULIRGs often show
signs of free-free absorption, which can decrease the outputs at
1.4GHz. To estimate this effect, we took the 19 star forming galaxies
with adequate data from \citet{condon91} and computed an average
slope for them of $-0.58$ between 1.49 and 8.44 GHz. Taking the 8.44 GHz
value to represent the un-absorbed radio output and assuming the
intrinsic slope is $-0.7$ \citep{condon91}, we revise the IR/radio
ratio downward to 140 to take account of absorption; this value is
calculated by extrapolating from 8.44 GHz to 1.4 GHz and comparing
results for slopes of $-0.58$ and $-0.7$.  The second concern is that the
radio-infrared ratio may be a function of redshift 
\citep{kovacs06,vlahakis07}. %(Kovacs et al. 2006; Vlahakis et al. 2007). 
We tested this suggestion by
determining a K-correction between 1.4 GHz and 850$\mu$m for an
average of local LIRG and ULIRG templates (to be published by
Alonso-Herrero et al.) and applying it to a large sample of high
redshift sources. We find that the K-correction accurately reproduces
the observed behavior with no offset, indicating there is no
significant evolution of the relation. We have therefore used a value
of 140 for the intrinsic flux density ratio at rest wavelengths to
convert the radio measurements to TIR ones. We extrapolated the radio
fluxes observed to a rest frequency of 1.4 GHz with a slope of
$-0.7$. The redshift usefully makes the observed rest frequency
significantly higher than 1.4 GHz, reducing the effects of free-free
absorption for the high-z galaxies. To convert to L(TIR), we use the
relationship from our local sample,

L(TIR)$ =$ 740 $\times $ L(60)$^{\alpha}$, where $\alpha = 0.9427$

\noindent with an rms scatter of 0.08 dex.  

It is important to point out that both of the corrections we have
derived have the effect of {\it reducing} the change we find in
L(8$\mu$m)/L(TIR). That is, using the relation derived by \citet{yun01}
and applying the proposed change at high redshift \citep{kovacs06,vlahakis07}
would have both resulted in a much larger change than reported. 
Therefore, we are taking a conservative approach.

We take  radio flux densities for the galaxies observed by \citet{yan07}
from \citet{condon03} as tabulated on the 
SSC website\footnote{http://ssc.spitzer.caltech.edu/fls/extragal/vla.html}.
% added this back in
Two of the remaining galaxies are
detected at low significance as indicated by inspection of the grey
scale images for them (MIPS 429 and 506), so we enter rough values of
0.1~mJy for them.
For MIPS 289, which is undetected, we assigned an upper limit of 0.1~mJy.
We reject the source MIPS 8493 because of fringing in the VLA image at its
position.  The other four (MIPS 289, 15928, 16144, and 22530) are in
regions where the radio images are clean.
We correct their  observed radio measurements to the rest
frame assuming a radio k-correction of $(1+z)^{0.7}$, since a typical
slope for a starburst radio spectrum is $-0.7$ \citep{condon91}. 
We plot the resulting L(TIR) and L(8~\micron) values in 
Figure~\ref{fig:newTIR} as circles (filled for the detections, 
open for the upper limit.)

%These three tests (using data from \citealt{yan07}, \citealt{papovich07}, and
%\citealt{daddi05}) all indicate 
%that the rest--frame 8~\micron\  outputs of galaxies at 
%$z \sim$ 2 are significantly brighter relative to TIR or submm fluxes than is
%typical of local templates.  However, in two of these cases L(TIR)
%exceeds that of all the local analogs; even the average L(TIR) of the
%fainter of the two bins of \citet{papovich07} is slightly above
%the most luminous local ULIRGs. For the samples of Daddi \etal\ 
%and Papovich \etal, AGNs have been screened out by requiring the absence
%of strong X-ray emission, or in the second case, by also eliminating
%objects with power-law IRAC SEDs. The discrepancy is therefore
%strongly implied, but not demonstrated completely because of both the
%disconnect in luminosity ranges and the lack of direct mid-IR spectroscopic
%confirmation that the rest 8~\micron\ flux is dominated by power from
%star formation, not from AGN.

\subsubsection{8~\micron-to-TIR ratio for the $z \sim 2$ lensed galaxies}

The lensed galaxies described in this paper bridge the luminosity divide
between the $z\sim 2$ extremely high--luminosity galaxies in the literature,
and the comparison LIRG and ULIRG samples at $z\sim 0$ and $z = 0.85$.
For the lensed sample we can determine which galaxies have mid-IR spectra
dominated by star formation, using the aromatic feature strengths.
For the subset of lensed galaxies with submm measurements and good upper
limits at 70~\micron, the SEDs are well enough constrained for a direct
determination of L(TIR).  Therefore, the lensed galaxies can cleanly test
for evolution with redshift in the behavior of the L(8~\micron)/L(TIR) ratio.

We use the aromatic-to-mid-IR flux ratio introduced in \S\ref{sec:nuc}
to select galaxies whose mid-IR spectra are dominated by star formation.
The ratio selects A2219a, A1689b, A2218a, and A1835a as all having strong 
aromatic emission; their ratios range from  7--24 $\times 10^{-11}$, 
with an average value of $16 \times 10^{-11}$.
By comparison, our sample of local star--forming LIRGs and ULIRGs has a range 
in ratio of 7--32 $\times 10^{-11}$, with an average of $21 \times 10^{-11}$.  
That is, the relative aromatic feature strength is virtually the same 
for the two samples, with no evidence that AGN have 
significantly augmented the 8~\micron\ fluxes of our star--forming 
high--z sub-sample.  
The EW(6.2~\micron) criterion we used for the \citet{yan07} sample
would have returned the same sample of star--forming lensed galaxies, 
except that A2219a lacks 6.2~\micron\ coverage and thus would not have
been selected.

We plot the four lensed galaxies as red triangles in Figure~\ref{fig:newTIR}.

% new, 20 june 2007
\subsection{Comparison of 8~$\mu$m-to-TIR ratio at high and low redshift}

The L(8~$\mu$m)/L(TIR) ratio is important for two reasons: 
1) it is one of the most accessible measures of the physical behavior of
infrared emission as a function of redshift; and 
2) many studies rely on fluxes at observed--frame 15 or 24~\micron\ 
 to deduce L(TIR) and related properties of galaxies at $z \sim 1$ or 2,
and are thus relying on the behavior of L(8~$\mu$m)/L(TIR). 
Figures~\ref{fig:TIRL8} and \ref{fig:newTIR}
illustrate the important result that this ratio is well--behaved 
for star--forming galaxies out to $z \sim 2.5$. 
In fact, this good behavior is a
reasonably demanding test, since the various samples at $z \sim 2$ are
subject to different selection effects and have L(TIR) derived in
different ways, yet they all follow a single trend. 
To recapitulate the different methods:
 the lensed galaxies are selected largely on amplification 
factor, and we measured L(TIR) directly from template fits to rest 
frame mid- and far-IR and submm photometry. 
The four galaxies from \citet{yan07}
were selected on the basis of colors and brightness at 24~\micron, and we
estimated L(TIR) from the radio-infrared relation.  The stacked
galaxies from \citet{papovich07} were selected based on faint
K-band and 24~\micron\ detections; we estimated L(TIR) from rest 24~$\mu$m
measurements.  The stacked galaxies from \citet{daddi05}  were
originally selected from BzK colors indicating they are at $z \sim 2$
and were stacked based on selections at a variety of wavelengths, and
we used rest submillimeter measurements to estimate L(TIR).

Having noted the overall uniform behavior, and the general similarity
of this behavior independent of redshift, we now evaluate the evidence
from Figures~\ref{fig:TIRL8} and \ref{fig:newTIR}   %%7 and 8 
for modest evolution with redshift. Specifically,
there is a tendency for the ratio L(8$\mu$m)/L(TIR) to decrease with
increasing luminosity for the local galaxies for L $> 10^{11}$
L$_\odot$, but this tendency seems reduced for the galaxies at
$z \sim  2$.  In fact, the high redshift galaxies appear to have SED
behavior similar to normal local galaxies of lower luminosity in this
regard.

We have made two tests of the significance of this difference. The
first is based on the stacking results of \citet{daddi05} and
\citet{papovich07}.
Both studies select the galaxies to stack on
the basis of K-band detections, and therefore the mid-- and
far--infrared characteristics of the galaxies should be representative
of the entire population of IR-active galaxies (i.e., there is no bias
toward a specific L(8$\mu$m)/L(TIR). To have an analogous set of data
for the local galaxies, we have averaged their L(8$\mu$m)/L(TIR)
ratios in luminosity intervals of $10^{0.5}$; these averages are shown
as the large blue triangles in Figure~\ref{fig:TIRL8}.  %figure 7. 
We have then fitted the averaged data points, shown as a blue line.  
To test whether the high
redshift stacked results were compatible with the same fit, we carried
out a second fit in which we included the three stacked points. For
the galaxies above $10^{11}$ L$_\odot$, we then compared $\chi^2$ {\it
for the low redshift galaxies only} in the two cases. We found that
this $\chi^2$ is 4.5 times larger if the three stacked points 
at $z\sim 2$ 
are included in the fit; that is, there is a strong indication that a
single functional fit cannot be used for both the local galaxies and
for those stacked at $z \sim 2$.\footnote{This analysis 
makes the assumption that low--luminosity galaxies have similar 
behavior at low and high redshift.}

%new Oct 24 2007
To determine whether a few anomalous local galaxies could affect this 
conclusion, we repeated the procedure but using a more robust average 
for the luminosity-binned ratios. For each bin with ten or more items, 
we discarded the two highest and two lowest ratios and then averaged the 
remainder. Where there were less than ten items, we retained the straight 
average. The highest luminosity bin, which is responsible for most of the 
change in slope of the relation between the two luminosities, has ten 
objects and hence its average was recomputed. The results were virtually 
identical to those with the straight averages, as shown in Figure 7.

A second test was applied to the high redshift galaxies measured
individually, either by \citet{yan07}, or by us. In the case of
the Yan \etal\ sample, we expect a strong bias toward objects with a
large ratio of L(8$\mu$m)/L(TIR), because they selected relatively
bright objects for spectroscopy, given the sensitivity limitations of
the IRS. Our sample was selected on a combination of 24$\mu$m flux
density and lensing amplification, so a bias is likely, although
perhaps less severe.

To compare with these two studies, we selected a local sample with
relatively high 8~$\mu$m output by removing all the galaxies that are
below the fit made to the averaged values (above). This
sample is shown in Figure~\ref{fig:newTIR}. %8. 
It is noteworthy that the scatter in L(8$\mu$m)/L(TIR) is small, 
with no outliers toward high L(8$\mu$m),
so this is an appropriate comparison sample for cases where a bias is
suspected. As before, we carried out two fits, one just to the local
galaxies, and another to them plus the eight individually measured
high redshift ones. We found that $\chi^2$ for the low redshift sample
with L $> 10^{11}$ L$_\odot$ increased by a factor of 2.5 
if we included the high redshift galaxies in the fit.  
Comparing Figure~\ref{fig:newTIR} 
with Figure~\ref{fig:TIRL8}, the discrepancy would
be substantially larger if we had included all the local galaxies
in our fits.   This result confirms that found for the
stacked samples, namely that a single functional fit is not a good fit
to the data over all redshifts. 

Another comparison can be made with the sample of infrared galaxies 
at $z \sim 0.85$ from \citet{delphine06}.
The results fall in a region occupied by both low and high
redshift fits in Figure~\ref{fig:newTIR}, % 8
which confirms the relatively modest change
in L(8$\mu$m)/L(TIR) with redshift, but (not surprisingly given the
small difference between $z \sim 0$ and 2) does not definitively
deviate from either fit. In making this comparison, we have assumed
that the galaxies studied by \citet{delphine06} have a bias
toward bright 15$\mu$m observed flux densities. If instead we treat
them as an unbiased sample and average them in luminosity intervals,
they fall along (or slightly above) the fit for the $z \sim 2$ sample
in Figure~\ref{fig:TIRL8}.  %7 

The tendency at high redshift for relatively strong 8$\mu$m emission 
could arise either through a contribution of an extra,
AGN-powered component at 8~$\mu$m 
(e.g. the extra 5.8~\micron\ continuum claimed by \citealt{sajina07}
and the mid-IR excess galaxies with hard X-ray emission discussed by
\citealt{daddi07b}), 
or to a change in the
properties of the stellar--powered infrared--emitting regions.  The AGN
explanation runs counter to our selection of individual galaxies with
strong aromatic features that correspond to equivalent widths typical
of purely star forming local galaxies. It is also counter to the
precautions taken in the stacking analyses to exclude AGN (e.g.,
\citealt{papovich07}).  
In addition, AGN contributions would come in a
broad range, so it would be hard to explain the consistent behavior of
the five high redshift samples in Figures~\ref{fig:TIRL8} and \ref{fig:newTIR}
if AGN were causing the departure from the behavior from local star 
forming galaxies. 
Finally, the behavior we see is similar to that found by \citet{xzheng07}
in a stacking analysis at $z \sim 0.7$; specifically, they report that
the far-IR SEDs appear to resemble those of local galaxies of lower
luminosity.  Thus, while AGN may boost 8~\micron\ emission in other sources, 
it is unlikely that AGN contamination is responsible for the behavior 
seen in Figures~\ref{fig:TIRL8} and \ref{fig:newTIR}.

We conclude that star-forming galaxies have a mild trend toward increasing 
L(8~\micron)$/$L(TIR) ratio with redshift that 
reaches up to a factor of $\sim 2$ at $z=2$.
In other words, L(8~\micron) does not turn over with increasing L(TIR) for $z\sim2$ 
galaxies as it does at low redshift.  This probably helps explain why some 
high--redshift galaxies have higher $6.2$\micron\ aromatic equivalent widths 
than are observed for local ULIRGs \citep{desai07,md07,sajina07},
%Menendez-Delmestre et al. 2007; Sajina et al. 2007), 
though direct comparisons cannot be made because there are not local galaxies 
with such extremely high luminosities.   Because the lensed galaxies and 
stacked samples directly probe the $10^{11}$--$10^{12.5}$~\Lsun\ range at 
$z\sim 2$, we can directly compare to $z\sim 0$ galaxies at similar luminosity 
rather than extrapolating.

That aromatic luminosities do not turn over with L(TIR) in high redshift galaxies
may be related to lower densities and optical depths of
the infrared--emitting regions in the high-z galaxies.  Lower optical
depths would result in part if the metallicity were lower, resulting
in a higher gas-to-dust ratio.  \citet{draine07} show that the dust 
content of galaxies has a very steep dependence on their metallicity. 
The densities and optical depths of the
star forming regions would also be lower if the high luminosities are
associated with star formation on large spatial scales in the high--redshift 
galaxies, as opposed to strongly--concentrated, major-merger--associated
nuclear star formation in local ULIRGs.
Such very dense regions in local ULIRGs appear to account for the scattering 
of points below the $z = 0$ trend line.

Use of local templates to estimate L(TIR) for high-z galaxies
detected at 24~$\mu$m has indicated the existence of very high
luminosities at $z \sim 2$ and beyond (e.g., \citealt{yan07}.) 
The change in L(8~\micron)/L(TIR) suggested here would reduce
L(TIR) for a given L(8~\micron) by a factor of $\sim 4$.
As shown in the figure, in our sample there is only one galaxy
selected to be dominated by star formation that has a L(TIR) significantly 
exceeding $10^{13}$~\Lsun.\footnote{This galaxy was selected as star--forming
based on its EW(6.2~\micron) \citep{sajina07}, but its EW(7.7~\micron) is 
unusually low, which may indicate a substantial AGN contribution.}
A luminosity of $10^{13}$~\Lsun\ corresponds to a star formation rate of 
$\sim 2000$~ \Msun\ yr$^{-1}$ if we use the local calibration \citep{kennicutt98},
and is suggestively close to the theoretical maximum luminosity that a
starburst is allowed by causality arguments---that star formation 
throughout a galaxy
should not transpire faster than the dynamical timescale \citep{lh96,elmegreen99}.
(The actual maximum value depends on efficiency and dynamical mass.)
Thus, our lower calibration for L(8~\micron)/L(TIR) may remove the discrepancy
between L(TIR) inferred for high redshift galaxies and plausible upper limits
on what a starburst can produce.  Spatially--resolved imaging of high--redshift 
starbursts (e.g. \citealt{nesvadba07}) can test whether they are in fact 
forming stars at the dynamical limit.

\section{Summary and Implications}

The 24~\micron\ photometric band of Spitzer efficiently detects
galaxies out to $z\sim 3$, and is being
used as an extinction--robust measure of
star formation rates in the distant universe.  
We present new IRS spectra and MIPS photometry of 
gravitationally--amplified galaxies, 
and use them in concert with archival Chandra images and published 
submillimeter photometry to investigate the nature of the very 
faint 24~\micron\ sources.

The aromatic feature flux ratios observed in our spectra agree well with
those of star--forming template spectra at $z\sim 0$.  This result
implies that dust size and
ionization distribution are not strongly evolving with redshift.
Thus, for $z \ga 1$ star--forming galaxies with little AGN
contribution, the mid-IR spectra can be approximated by $z\sim 0$
star--forming templates.

At $z \sim 2$, the 24~\micron\ data probe the 8~\micron\ rest wavelength region. 
We find that the ratio of rest-frame 8~\micron\ luminosity to total infrared
luminosity is approximately the same as for local galaxies of similar L(TIR). 
However, the $z \sim 0$ relationship does not appear to hold perfectly.
Combining data from the literature with
this sample, we demonstrate that $z \approx 2$ star--forming galaxies
have higher L(8~\micron) luminosities for a given L(TIR) than
low--redshift analogs.  Functionally, the break in the 
L(8~\micron) vs. L(TIR) relationship occurs at a 10-times-higher L(TIR) at 
$z\sim2$ than it does at $z\sim 0$.
We speculate that this is due to
lower optical depths in the IR--emitting regions of the high--z
galaxies, perhaps because of lower metallicity or because
star formation is more spatially extended
than in local ULIRGs (e.g., across a disk rather than in a compact
nucleus).  This result has important implications for mid-IR--derived
star formation rates at $z\sim 2$, and may indicate a maximum
star--forming luminosity of a few  $\times 10^{13}$~\Lsun.

The results just summarized were found for sources where AGN do not
contribute substantially to the mid-IR spectra, as determined by
the aromatic-to-mid-IR flux ratio.  But 6 of the 19
mid-IR spectra in our extended sample {\bf do} show a strong AGN.
Including other AGN diagnostics (X-ray
detections and optical emission lines), 8 of the 19 galaxies show
evidence for nuclear activity.  The prevalence of AGN in this
sample may contradict the frequent assumption that the 24~\micron\
flux density in 100~\microJy\  sources at $1<z<3$ is dominated by star
formation, with AGN a modest contaminant.

\acknowledgments
We thank J.~Donley for computing X-ray flux upper limits 
for the non-detections, 
and D.~Sand for making available his reduction of the WFPC2 images.
We thank D.~Lutz and H.~Teplitz for sending  
electronic versions of their published spectra.
This work is based in part on observations made with the Spitzer
Space Telescope, which is operated by the Jet Propulsion Laboratory,
California Institute of Technology under a contract with NASA. Support
for this work was provided in part by NASA through contract 1255094 
issued to the University of Arizona by JPL/Caltech. 
J.~Rigby was supported in part by NASA through the
Spitzer Space Telescope Fellowship Program.
This work has made use of archival data from the Chandra X-ray Observatory,
the Hubble Space Telescope, and the Spitzer Space Telescope.
This research has made use of NASA's Astrophysics Data System Bibliographic
Services and the NASA/IPAC Extragalactic Database.

\appendix
\section{Notes on Individual Sources}

\subsection{Source A2390a}  %eiichi's notation is 2390a
%This source was discovered in the 24~\micron\ image  (\fnu $= 0.83$~mJy),
%and has a faint, irregular HST counterpart.
Johan Richard obtained a Keck spectrum for this source,
shown in  Figure~\ref{fig:johan_spec}.  
The strong Lyman $\alpha$ emission is redshifted by $z=2.858$,
and the line velocity spread (FWHM $\sim 945$ km s$^{-1}$) 
indicates that the source is an AGN.

%\subsection{Source Abell 2390b}  %eiichi's notation is 2390b
%This source's HST morphology is complex, and it has a spectroscopic 
%redshift of 0.913 \citep{pello91}.   
%It was also detected by ISO \citep{lemonon98}.

\subsection{Source A2261a}
This sub-mm galaxy \citep{chapman02}
lacks a spectroscopic redshift; submillimeter and radio-based  photometric 
estimates place it  at $z>1.6$ \cite{chapman02} and $2.5<z<4$  \citep{aretxaga03}.
It is the only source without a redshift in our sample.
Our spectrum should be sensitive to the 
11~\micron\ complex for $0.9<z<1.8$, and to the
7.7~\micron\ aromatic feature for $1.8<z<3.1$.  
Thus, for the aromatic features to fall 
outside the IRS spectral coverage, the source would have to have 
very low ($z<1$) or very high ($z>3.1$) redshift.    If the source were at 
$z>3.1$, it must be $\sim 1000$ times more luminous than M82 to have produced
the observed 24~\micron\ flux density.  It seems more probable that 
this source lies at $1<z<3$ and lacks aromatic emission.

%\subsection{Source A851a}
%Three PAH features set the source's redshift as $z=2.38$.  Two 
%additional, narrow emission lines are present at 15.0 and 17.1~\micron,
%which we tentatively identify as 
%H$_2$ S(10) 4.41~\micron\ and H$_2$ S(8) 5.05~\micron.

%%%%%%%%%%%%%%%%%%%%%%%%%%%%%%%

\clearpage

\begin{deluxetable}{lllllllll}
%\rotate (temp)
\tabletypesize{\scriptsize}
\tablecolumns{9}
\tablewidth{0pc}
\tablenum{1}
\label{tab:sample}
\tablecaption{The IRS lensed sample.}
\tablehead{
\colhead{} & \colhead{RA(J2000)} &  \colhead{DEC(J2000)} & \colhead{sub-mm name} & \colhead{ref} & 
\colhead{PID} & \colhead{z} &  \colhead{ref} & \colhead{amp}\\}
\startdata
\cutinhead{Our sample}
%#src     RA(J2000)    DEC(J2000)     submm-name        ref  PID   z     ref  amp    
MS0451a & 04:54:07.12 & $-$03:00:36.4 &                &   & 1 & $1.95$  &1&  $2.5$       \\ 
A851a   & 09:42:54.55 & +46:58:44.3   &SMM J09429+4658 & A & 2 & $2.38$  &1&  $1.3,2.0$   \\ 
A1689a  & 13:11:28.04 & $-$01:19:18.7 &                &   & 2 & $1.1537$&2&  $8.5\pm2.3$ \\ 
A1689b  & 13:11:29.14 & $-$01:20:46.5 &SMM J13115-01208& B & 2 & $2.63$  &2&  25--47      \\ 
A1835a  & 14:01:04.96 & +02:52:24.8 &SMM J140105+025223.5 & C &12 & $2.565$ &3&  $3.5\pm0.5\tablenotemark{a}$ \\
A2218a  & 16:35:54.18 & +66:12:24.8   &  \imageB       & D & 1 & $2.516$ &4&  $22\pm2$    \\ 
A2218b  & 16:35:55.16 & +66:11:50.8 &SMM J163555.2+661150&E & 2 & $1.034$ &5&  $6.1$      \\ 
A2218c  & 16:35:59.12 & +66:11:47.4   &                &   & 2 & $0.97$  &1& 6.7          \\ 
A2219a  & 16:40:19.50 & +46:44:00.5   &SMM J16403+46440 & F & 1 & $2.03$  &1& $3.6$       \\ 
A2261a  & 17:22:21.00 & +32:07:05.5   &SMM J17223+3207  & F & 12& $1<z<3$ &1& $3.3$       \\ 
A2390a  & 21:53:33.24 & +17:42:10.6   &                &   & 12& $2.858$ &6& $10.8$       \\ 
A2390b  & 21:53:34.43 & +17:42:21.6   &                &   & 12& $0.913$ &7& $3.35$       \\
A2390c  & 21:53:34.540 & +17:42:03.0  &`Source 14'     & G &  2& $0.91$  &8& $10.4$       \\ %%% late-add
AC114a  & 22:58:49.79 & $-$34:48:47.0 &                &   &  2& $1.47$  &1&  $9.7$       \\
A2667a  & 23:51:40.00 & $-$26:04:52.0 &                &   &  2& $1.034$ &9& $17$         \\ %%% late-add
\cutinhead{Additional sources from the literature}
Teplitz-1    & 3:32:44.00 & -27:46:35.0 &              &   &  8 & 2.69   &10& 1.0    \\  %Teplitz 
Teplitz-1-BzK& 3:32:38.52 & -27:46:33.5 &              &   &  8 & 2.55   &10& 1.0    \\  %Teplitz 
Teplitz-2    & 3:32:34.85 & -27:46:40.4 &              &   &  8 & 1.09   &10& 1.0    \\  %Teplitz 
%2-x         & 3:32:39.72 & -27:46:11.3 &              &   &  8 & 1.55?  &10& 1.0 &  $0.225$ \\  %Teplitz 
%A370a    & 39.966125 & $-$1.59966 &SMM J02399-0136 & A &  9 & 2.81   &11& 2.45 & $1.36$ \\ % Lutz   SMM J02399-0136 \citep{smail02mnras}.}
A370a    & 2:39:51.87 &  -1:35:58.78  & SMM J02399-0136 & A &  9 & 2.81   &11& 2.45  \\ % Lutz   SMM J02399-0136 \citep{smail02mnras}.}
%Abell 2125  & 235.36366 & 66.27138  & 9 & 1 & 2.80  &lutz&      & $0.07$\tablenotemark{c}\\ 
\enddata
\tablecomments{Columns:\\
1) source;\\
2--3)  coordinates;\\
4) sub-mm name, if source has a sub-mm detection in the literature;\\
5) reference for detected sub-mm counterpart:
``A'' is  \citet{smail02mnras};	  
``B'' is  \citet{knudsenthesis} and \citet{knudsen07};	  
``C'' is  \citet{smail00,ivison00};  
``D'' is  \citet{kneib04};		  
``E'' is  \citet{knudsen06};	  
``F'' is  \citet{chapman02};	  
``G'' is  \citet{cowie02}.\\
6) Spitzer program in which IRS spectra were obtained:
``1'' signifies PID 82;  ``2'' signifies PID 30775; ``8'' signifies PID 252;
``9'' signifies PID 3241;\\
7) redshift;\\
8) redshift reference: 
``1'' means  redshift was determined from our IRS spectra;
``2'' is based on Keck spectra (J. Richard \etal\ in prep.);
``3'' is from \citet{frayer99};  
``4'' is from \citet{kneib04};
``5'' is from \citet{ebbels98}; 
``6'' is from  Keck spectra presented in the Appendix;
``7'' is from \citet{pello91}; 
``8'' is from \citet{frye98}; 
``9'' is from \citet{sand05}; 
``10'' is from \citet{teplitz07}; 
``11'' is from \citet{frayer98} and also \citet{lutz05}.\\
9) amplification estimated from the lensing models.  Values from the 
   literature are footnoted; un-footnoted values are new;\\ %from Kneib and Richard;\\
%10) flux density at 24~\micron, from our DAOPHOT PSF-fitting photometry 
%on images from Spitzer PID 82;\\
%11) total IR luminosity derived from template fitting, in \Lsun :  best-fit is listed first, 
%and then the range of plausible fits;\\
%12) template used for L(TIR) fit.\\
}
\tablenotetext{a}{The amplification for this source has been the subject of several papers;
\citet{downessolomon} and \citet{motohara} argue that 
an intervening cluster galaxy boosts the amplification to $\sim 25$.  
\citet{smail05} and \citet{gsmith05} find that the intervening galaxy is a dwarf 
with small velocity dispersion and thus contributes only modestly to the total 
amplification, which they estimate as $3.5\pm0.5$.  We note the controversy 
and adopt the \citet{smail05} amplification.}
%\tablenotetext{b}{This is source SMM J02399-0136 \citep{smail02mnras}.}
%\tablenotetext{c}{Tentative detection at 22~\micron\ from \citet{charman04} with IRS peak-up imaging.}

\end{deluxetable}
\begin{deluxetable}{lllll}
%\rotate
%\tabletypesize{\small}
\tablecolumns{5}
\tablewidth{0pc}
\tablenum{2}
\tablecaption{Exposure times.\label{tab:exptime}}
\tablehead{
\colhead{Cluster} &  \colhead{t(24~\micron)} &  \colhead{t(3.6~\micron)} & \colhead{t(IRS, LL)} & \colhead{t(Chandra)} \\
}
\startdata
%clust$  &  24tot  &  tch1   &  IRS   \\
MS0451  & $3.68$  &  $2.40$  &  3.66\tablenotemark{b}        & 53\\
A851   &  $2.78$  &  $2.40$  &  6.34\tablenotemark{a}        & 0\\
A1689  &  $2.78$  &  $2.40$  &                               & 40 \\
 ~~~a  &          &          &  7.31\tablenotemark{a}        & \\
 ~~~b  &          &          &  14.63\tablenotemark{b}       & \\
A1835  &  $2.77$  &  $6.00$  &  3.66\tablenotemark{b}; 3.66\tablenotemark{a}  & 29  \\ %(IRAC from IRAC-GTO, now public)  
A2218  &  $2.78$  &  $2.40$  &                               & 58\\
 ~~~a  &          &          &  3.66\tablenotemark{b}        & \\
 ~~~b  &          &          &  3.66\tablenotemark{a}        &\\
 ~~~c  &          &          &  7.31\tablenotemark{a}        & \\
A2219  &  $2.78$  &  $2.40$  &  3.66\tablenotemark{b}        & 41\\
A2261  &  $2.77$  &  $2.40$  &  5.85\tablenotemark{b}; 7.31\tablenotemark{a}  & 33\\
A2390  &  $2.77$  &  $2.40$  &                               & 109\\
 ~~~a  &          &          &  3.66\tablenotemark{b}; 5.61\tablenotemark{a}  & \\
 ~~~b  &          &          &  3.66\tablenotemark{b}; 5.61\tablenotemark{a}  & \\
 ~~~c  &          &          &  7.31\tablenotemark{a}        & \\
AC114  &  $2.77$  &  $2.40$  &  7.31\tablenotemark{a}        & 70\\
A2667  &  $2.77$  &  $2.40$  &  1.95\tablenotemark{a}        & 9.2\\

\enddata
\tablenotetext{a}{Both LL1 and LL2 were given this exposure time.}
\tablenotetext{b}{Only order LL1.}  %20-38 um
\tablecomments{Exposure times, in kiloseconds, for the 
following bands:  Spitzer/MIPS~24~\micron;
 Spitzer/IRAC 3.6~\micron; and Spitzer/IRS long-low grating.   
The 24~\micron\ exposure time is the median (per-pixel) exposure 
time within the central 4\arcmin\ by 4\arcmin\ box.}
\end{deluxetable}
%%
%
%note, 24um exptime found by multiplying the exptime/DCE by  the per-pixel median number of DCEs found within the 
%4'x4' central box. (from [noverlap]).  Outskirts are shallower, very central strip is deeper.
%
% IRS texp  is the exposure time given by spot in the ``resource estimates''
%box, (``Calc. Obs. Time), box irs_lo_20(sec)

  %\input{exptime}
\begin{deluxetable}{llllllllll}
%\rotate
\tablecolumns{10}
\tablewidth{0pc}
\tabletypesize{\small}
\tablenum{3}
\label{tab:longwave}
\tablecaption{Long-wavelength photometry and template fits.}
\tablehead{
\colhead{} & \colhead{f(3.5~\micron)} &  \colhead{f(4.5~\micron)} & \colhead{f(5.7~\micron)} & \colhead{f(8.0~\micron)} &
\colhead{f(24~\micron)} &  \colhead{(PAH/tot)} 
& \colhead{f(70~\micron)} & \colhead{L(TIR)} & \colhead{template}}
\startdata
% src         ch1        ch2         ch3        ch4       f24(mJy)  pah/total f(70,mJy)         L(TIR)  
MS0451a &  $32\pm2$ &  $49\pm2$ &  $52\pm3$ &  $36\pm3$ & $1.32$ &  6.3E-11 & $5.7 \pm 1.6$  & $<2.2 \times 10^{13}$             & CE \tablenotemark{a}\\  
A851a   &  $29\pm3$ &  $44\pm3$ &  $58\pm3$ &  $45\pm4$ & $0.63$ &  8.4E-11 & \nodata	     &  2.7 (2.3--3.6) $\times 10^{13}$\tablenotemark{b} & DH \\ %new fit
A1689a  &  $36\pm2$ &  $35\pm2$ &  $28\pm2$ &  $66\pm4$ & $0.58$ &          & \nodata	     &  \nodata                          &   \\
A1689b  &           &           &           &           & $0.32$ &  2.4E-10 & \nodata	     &  1.6 (0.43--2.4) $\times 10^{11}$ & DH\\
A1835a  &  $88\pm2$ & $105\pm2$ & $136\pm3$ & $124\pm5$ & $0.99$ &  1.4E-10 &  $<13$	     &  6.8 (4.2--11)   $\times 10^{12}$ & DH\\ 
A2218a  &  $67\pm2$ &  $70\pm2$ &  $92\pm6$ & $105\pm10$& $1.16$ &  2.0E-10 &  $<7$	     &  7.6 (5.7--9.5)  $\times 10^{11}$ & DH\\
A2218b  &  $182\pm4$& $147\pm3$ & $107\pm3$ & $107\pm5$ & $1.67$ &           & $7.4 \pm 1.5$  &  4.8 (2.8--5.3)  $\times 10^{11}$ & DH\\
A2218c  &  $99\pm3$ & $79\pm3$  & $67\pm2$  &$523\pm11$ & $0.50$ &          & $10 \pm 1$     &  1.3 (0.88--2.0) $\times 10^{11}$ & Mrk 231 \\ %new fit
A2219a  &           &           &           &           & $0.82$ &  7.1E-11 & \nodata	     &  2.9 (2.7--3.4)  $\times 10^{12}$ & DH\\
A2261a  &  $142\pm4$& $138\pm4$ & $117\pm3$ & $156\pm6$ & $0.58$ &          & $<3.2$	      &  ---\tablenotemark{c}             &   \\
A2390a  &  $39\pm3$ & $59\pm3$  & $90\pm3$  & $196\pm7$ & $0.83$ &  5.8E-11 & $<7.6$	     &  \nodata                          &   \\
A2390b  & $462\pm8$ & $334\pm6$ & $258\pm5$ & $199\pm7$ & $0.88$ &          & $5.3 \pm 1.6$  &  1.7 (1.1--2.5)  $\times 10^{11}$ & Mrk 231 \\ %new fit
A2390c  &           &           &           &           & $0.60$ &          & $<8$	     &  8.2 (5.5--11)   $\times 10^{10}$ & DH\\ % new fit
AC114a  &  $91\pm3$ & $97\pm3$  & $78\pm3$  & $76\pm4$  & $0.41$ &  1.4E-10 & \nodata	      &  \nodata                          &   \\
A2667a  & $466\pm6$ & $349\pm5$ & $269\pm4$ & $320\pm8$ & $1.52$ &          & \nodata	      &  \nodata                          &   \\
\cutinhead{Literature sample}
Teplitz-1 &         &           &           &           & $0.13$ & \nodata & & \\
Teplitz-1-BzK &     &           &           &           & $0.20$ & \nodata & & \\
Teplitz-2     &     &           &           &           & $0.13$ & \nodata & & \\
A370a         &     &           &           &           & $1.36$ & \nodata & & \\
\enddata
\tablecomments{Columns:
1) Source name.
2--5) IRAC photometry, in \microJy.  Formal errors from the photometry are quoted, though field
crowding probably and aperture corrections probably introduce additional $5$--$10\%$ errors.
The photometry for A2218a is superior since it is multiply imaged (with different crowding).
Crowding is too severe to quote photometry for A1689b, A2219a, and A2390c.
6) Observed 24~\micron\ flux density, in mJy, from DAOPHOT PSF-fitting. 
Errorbars from DAOPHOT are overly optimistic and not quoted; 
we use errorbars of 0.1~mJy for template fitting.
7) Aromatic--to--mid-IR flux ratio, as defined in \S\ref{sec:nuc}, for the $z>1.4$ sample.
8) Observed 70~\micron\ flux density, in mJy, from aperture photometry.
Errorbars are dominated by the sky.
9) Total infrared luminosity, our best--fit and the 
range of acceptable templates, in \Lsun.
10) Best-fitting template:  \citet{charyelbaz} or \citet{dalehelou}.
}
\tablenotetext{a}{Mrk 231 template may also fit.}
\tablenotetext{b}{Using amplification factor of 1.3.}
\tablenotetext{c}{L(TIR) cannot be usefully constrained because the redshift is unknown.}
\end{deluxetable}
  % long-wave photometry
\begin{deluxetable}{lllllllllllllllllllllll}
%\rotate
\tabletypesize{\scriptsize}
\tablecolumns{23}
\tablewidth{0pc}
\tablenum{4}
\tablecaption{Measured Aromatic fluxes.\label{tab:pahfluxes}}
\tablehead{
\colhead{src} &   \colhead{6.2} &  \colhead{$\sigma$} &  \colhead{7.7} &  
\colhead{$\sigma$} &  \colhead{8.3} &  \colhead{$\sigma$} &  
\colhead{8.6} &  \colhead{$\sigma$} &  \colhead{11.3} &  
\colhead{$\sigma$} &  \colhead{12.0} &  \colhead{$\sigma$} &  
\colhead{12.6} &  \colhead{$\sigma$} &  \colhead{13.6} &  
\colhead{$\sigma$} &  \colhead{14.2} &  \colhead{$\sigma$} &  
\colhead{16.4} &  \colhead{$\sigma$} &  \colhead{17} &  
\colhead{$\sigma$}
}
\startdata
%# fluxes and uncertainties in 10-15 erg/s/cm^2.
%#src    6.2   o 7.7  o   8.3  o    8.6um o   11.3 o   12.0 o   12.6  o  13.6  o  14.2  o   16.4 o  17  o 
MS0451a &  &  & $28$ & $6$ & $4.8$ & $2$ & &  & $4.3$ & $2$ &  &  & $5.0$ & $2$ &  &  &  &  &  &  &  & \\
A851a   & 2.8 & 0.6  & 15.6  & 0.8   & 2.4 & 1.  & 2.7 & 0.8   &  &    &     &       &     &  &  &  &  &  &  &  & \\
A1689a &&   &   $12.$&$5.$  &&  &   &   &  &  & &&  &  & &  & &&  &&  & \\  %right z=1.15
A1689b  & $6.5$ & $1.1$ & $21$ & $3$ & $3.8$ & $1.$ & $4.0$ & $2$ &  &  &  &  &  &  &  &  &  &  &  &  &  & \\
A1835a  & $13$ & $2$ & $39$ & $10$ & $7.3$ & $2$ & $7.8$ & $2$ &  &  &  &  &  &  &  &  &  &  &  &  &  & \\
A2218a  & $18.3$ & $0.9$ & $66$ & $3$ & $8.2$ & $1.$ & $13.5$ & $1$ &  &  &  &  &  &  &  &  &  &  &  &  &  & \\
A2218b  &  &  & $65$ & $3$ & $4.8$ & $1.7$ & $16.3$ & $2$ & $20$ & $3$ & $3.8$ & $1.2$ & $10.8$ & $1.3$ &  &  &  &  & $1.7$ & $0.8$ & $14$ & $5$  \\
A2218c  &  &  & $14.9$ & $0.9$ &  &  & $1.5$ & $0.8$ & $1.3$ & $0.4$ &  &  &  &  &  &  &  &  &  &  &  & \\
A2219a  &  &  & $19.3$ & $1.3$ &  &  & $4.6$ & $1$ & $5.6$ & $2$ &  &  &  &  &  &  &  &  &  &  &  & \\
A2390a  & $1.1$ & $1$ & $12.4$ & $3$ &  &  & $5.8$ & $2$ &  &  &  &  &  &  &  &  &  &  &  &  &  & \\ 
A2390b  &  &  &  &  &  &  &  &  & $14$ & $10$ &  &  & $6.1$ & $2.3$ &  &  &  &  &  &  &  &   \\
A2390c &       &       & 35  &  11  &   &    & 11.   &  8  & 17.  &  12  &   &  & 6.3   & 1.5 & & &       &     &  &  &  & \\
AC114a & $6.5$ & $0.5$ & $23$ & $2$ &  &  & $7.1$ & $1$ & $9.2$ & $1$ &  &  & $4.2$ & $2$ & & &  &  &  &  &  & \\
A2667a &       &       & 102  & 11  &  12   & 3     & 22    &  4  &  29   &  7  & 9.1   &  3  &  16   & 3   &  & &   &  &  5.6 & 1.9 &  &  \\

\enddata
\tablecomments{Measured aromatic dust emission feature fluxes and 
uncertainties, measured with PAHFIT.  
Units are $10 ^{-15}$~\cgsflux.    
Columnn headers give the central wavelength, in \micron, of each feature.
For features comprised of multiple components (like 7.7~\micron), 
total flux and uncertainty were calculated using PAHFIT's 
``main feature'' function.
}
\end{deluxetable}
\begin{deluxetable}{lrr}
%\rotate
%\tabletypesize{\small}
\tablecolumns{3}
\tablewidth{0pc}
\tablenum{5}
\tablecaption{X-ray fluxes and $3\sigma$ upper limits.\label{tab:xray}}
\tablehead{
\colhead{src} &  \colhead{f(2--8 keV)} &  \colhead{f(0.5--8 keV)} \\
              &   \cgsflux & \cgsflux
}
\startdata
%#src     f(hardX)  f(fullX)  Xdet? 
MS0451a & $< 1.55 \times 10^{-15}$ & $< 1.30 \times 10^{-15}$  \\
A851a   & \nodata                  &  \nodata \\
A1689a  & $< 7.9 \times 10^{-15}$ & $< 4.5 \times 10^{-15}$  \\
A1689b  & $< 1.1 \times 10^{-14}$ & $< 7.2 \times 10^{-15}$  \\
A1835a  & $< 5.8 \times 10^{-15}$ & $< 4.4 \times 10^{-15}$  \\
A2218a  & $< 5.2 \times 10^{-15}$ & $< 2.6 \times 10^{-15}$  \\ 
A2218b  & $< 2.8 \times 10^{-15}$ & $< 2.1 \times 10^{-15}$  \\ 
A2218c  & $< 2.4 \times 10^{-15}$ & $< 1.6 \times 10^{-15}$  \\ 
A2219a  & $< 3.4 \times 10^{-15}$ & $< 3.0 \times 10^{-15}$  \\ 
A2261a  &   $3.7 \times 10^{-15}$  &  $1.8 \times 10^{-15}$   \\
A2390a  &   $5.1 \times 10^{-14}$  &  $7.6 \times 10^{-14}$   \\
A2390b  &   $4.4 \times 10^{-14}$  &  $2.5 \times 10^{-14}$   \\
A2390c  & $< 7.5 \times 10^{-15}$ & $< 5.2 \times 10^{-15}$ \\
AC114a  & $< 2.8 \times 10^{-15}$ & $< 2.9 \times 10^{-15}$  \\
A2667a  & $< 3.2 \times 10^{-14}$ & $< 4.1 \times 10^{-14}$ \\
%# X-ray limits are Jenn's "3 sig lim+src":  the 3 sigma sky background 
%#  plus any additional  counts at the source position.
\enddata
\tablecomments{X-ray limits are the $3\sigma$ sky background plus 
any additional counts at the source position, as described in 
\S\ref{sect:xraydata}.
}
\end{deluxetable}
\begin{deluxetable}{llll}
\tablecolumns{4}
\tablewidth{0pc}
\tablenum{6}
\tablecaption{Aromatic ratios for \imageB\label{tab:pahrats}}
\tablehead{\colhead{lines} & \colhead{ratio} & \colhead{SINGS} & \colhead{Starburst} }
\startdata
%  wave   kind   flux        smith
7.7\micron/6.2\micron\ & $3.6 \pm 0.2$   &  3.6 (1.3--4.8)  &  3.9 \\ 
7.7\micron/8.6\micron\ & $4.9 \pm 0.4$   &  5.7 (4.7--9.0)  &  5.1 \\
6.2\micron/8.6\micron\ & $1.36 \pm 0.1$  &  1.5 (1.2--3.0)  &  1.3 \\
\enddata
\tablecomments{Columns: (1) line wavelengths (\micron); 
(2) measured flux ratios; (3) median flux ratios and
$10\%$--$90\%$ range of variation from the SINGS sample \citep{jdsmith07}; 
(4) flux ratios of the average starburst template of \citet{brandl06}.}
\end{deluxetable}

\begin{deluxetable}{lcccccc}
\tablecolumns{7}
\tablenum{7}
\tablewidth{0pc}
\tablecaption{[Ne III] 15.5~\micron\ and [Ne II] 12.8~\micron.\label{tab:ne3ne2}}
\tablehead{
\colhead{src} & \colhead{[Ne II]} & \colhead{$\sigma$} & 
\colhead{[Ne III]} & \colhead{$\sigma$} & 
\colhead{[Ne III]/[Ne II]} & \colhead{$\sigma$}}
\startdata
%src    &   NeII 12.8   &  dNeII	   & NeIII15.5    &  dNeIII    & 3/2  &	 drat\\
MS0451 &      --       & 	   & \nodata	  &            &      &      \\ 
A851   &  \nodata      & 	   & \nodata	  &            &      &      \\
A1689a &  0.93         & 0.26      & 2.0	  &  0.43      & 2.15 &  0.76\\
A1689b &  \nodata      & 	   & \nodata	  &            &      &      \\
A2218a &  \nodata      & 	   & \nodata	  &            &      &      \\
A2218b &  2.9          & 0.6       & 0.91	  &  0.31      & 0.31 &  0.13\\
A2218c &      --       &           &     --       &	       &      &      \\   
A2219  &  \nodata      & 	   & \nodata	  &            &      &      \\
A2390a &  \nodata      & 	   & \nodata	  &            &      &      \\
A2390b &  2.2          & 0.57      & 0.90  	  &  0.69      & 0.41 &  0.33\\
A2390c &  0.83         & 0.47      &     --       &	       &      &      \\   
AC114a &  1.6          & 0.29      & \nodata	  &            &      &      \\
A1835a &  \nodata      & 	   & \nodata	  &            &      &      \\
A2261a &  \nodata      & 	   & \nodata	  &            &      &      \\
A2667a &  2.5          & 0.49      & 0.92  	  &  0.51      & 0.37 &  0.22\\
Tep1   &  \nodata      & 	   & \nodata	  &            &      &      \\
T1-bzk &  \nodata      & 	   & \nodata	  &            &      &      \\
Tep 2  &  \nodata      & 	   & \nodata	  &            &      &      \\
A370a  &  \nodata      & 	   & \nodata	  &            &      &      \\
\enddata 
\tablecomments{Measured fine structure line fluxes,
from PAHFIT, in units of $10 ^{-15}$~\cgsflux.  
``\nodata'' means the wavelength regime was not covered; 
 ``--'' means the line was covered but not detected.
}
\end{deluxetable}
  % table of NeIII, NeII 
\begin{deluxetable}{lllll}
\tablecolumns{5}
\tablenum{8}
%\tablewidth{0pc}
\tablecaption{AGN Diagnostics.\label{tab:agn-diag}}
\tablehead{
\colhead{src} & \colhead{X} & \colhead{weak aromatic} & \colhead{rising} & \colhead{opt}}
\startdata
%1      &  2   3   4   5 \\
MS0451a &    & X & \nodata     & \nodata  \\ 
A851a   &\nodata &  & \nodata  & \nodata  \\ 
A1689a  &    & X & X           & X \\
A1689b  &    & 	 & \nodata     &   \\
A1835a  &    & 	 & \nodata     &   \\
A2218a  &    & 	 & \nodata     &   \\
A2218b  &    & 	 &             &   \\
A2218c  &    & X & X           & \nodata  \\
A2219a  &    &   & \nodata     & \nodata  \\
A2261a  &  X & X & X           & \nodata  \\
A2390a  &  X & X & \nodata     & X \\
A2390b  &  X &   &             &   \\
A2390c  &    & 	 &             &   \\
AC114a  &    & 	 &             & \nodata  \\
A2667a  &    & 	 &             &   \\
Tep-1   &  X & 	 & \nodata     &   \\
Tep-1-BzK&   &   & \nodata     & \nodata  \\
Tep-2   &    & 	 &             &   \\
%Tep-2-x &  X & - &            & \nodata  \\
A370a   &  X & X & X \nodata   & X \\
%A212   &    & - &             &   \\
\enddata 
\tablecomments{Columns:
1) source name
2) X-ray detection;
3) low aromatic flux contribution, as defined in \S\ref{sec:nuc};
4) mid-IR spectrum rises strongly toward the red;
5) AGN lines in optical spectrum.
}
\end{deluxetable}
\clearpage

%begin the figures here
%%%%%%%%%%%%%%%%%%%%%

\clearpage
\centerline{\includegraphics[width=6in,angle=0]{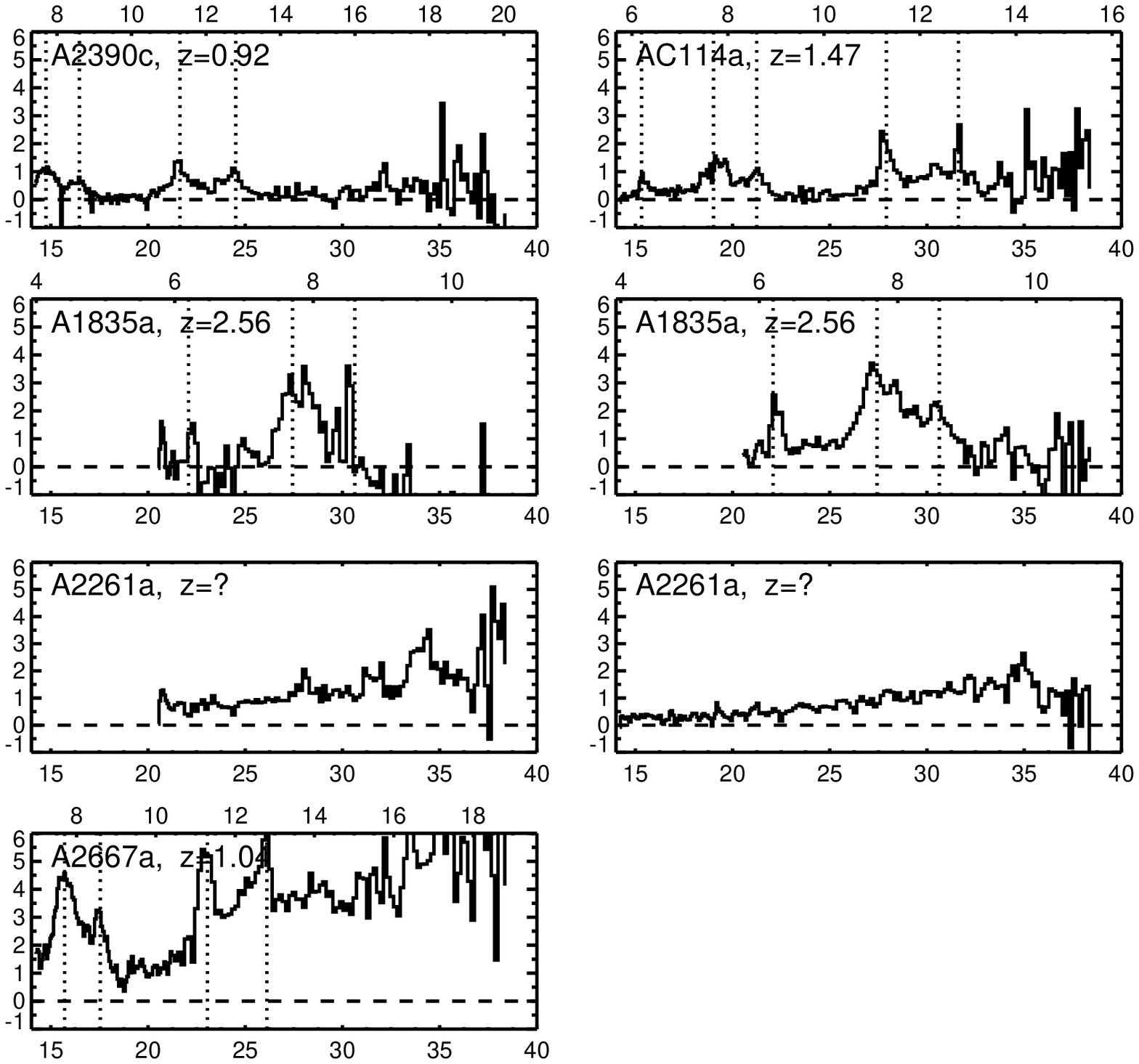}}
%{/WORK/Clusters/IRS_spectra_lensed/Plots/Eiichi_spec/p2.ps}
\centerline{Fig. 2 --- continued.}
\clearpage

%%%%%%%%%%%%%%%%%%%%%

%%%%%%%%%%%%%%%%%%%%%

%%%%%%%%%%%%%%%%%%%%%

%%%%%%%%%%%%%%%%%%%%%

\begin{figure}
\figurenum{6}
%\epsscale{0.5}
%\plottwo{f6b.epsi}{f6c.epsi}
%\plottwo{f6d.ps}{f6e.ps}
%\plottwo{f6f.ps}{f6g.ps}
%\plottwo{f6h.ps}{f6i.ps}
%\plottwo{f6j.ps}{f6k.ps}
%\epsscale{1.0}
\includegraphics[width=1.85in,angle=270]{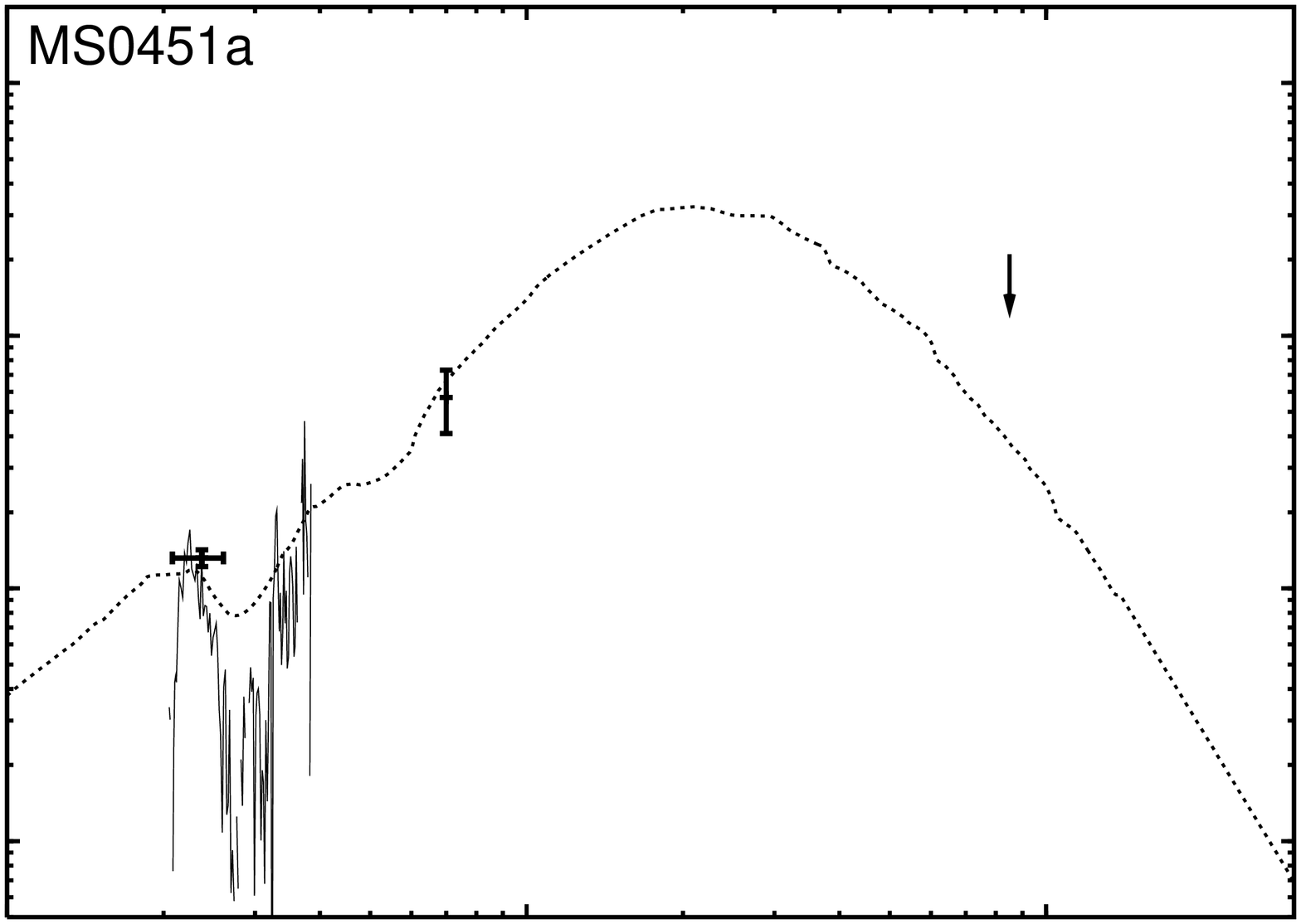}
\includegraphics[width=1.85in,angle=270]{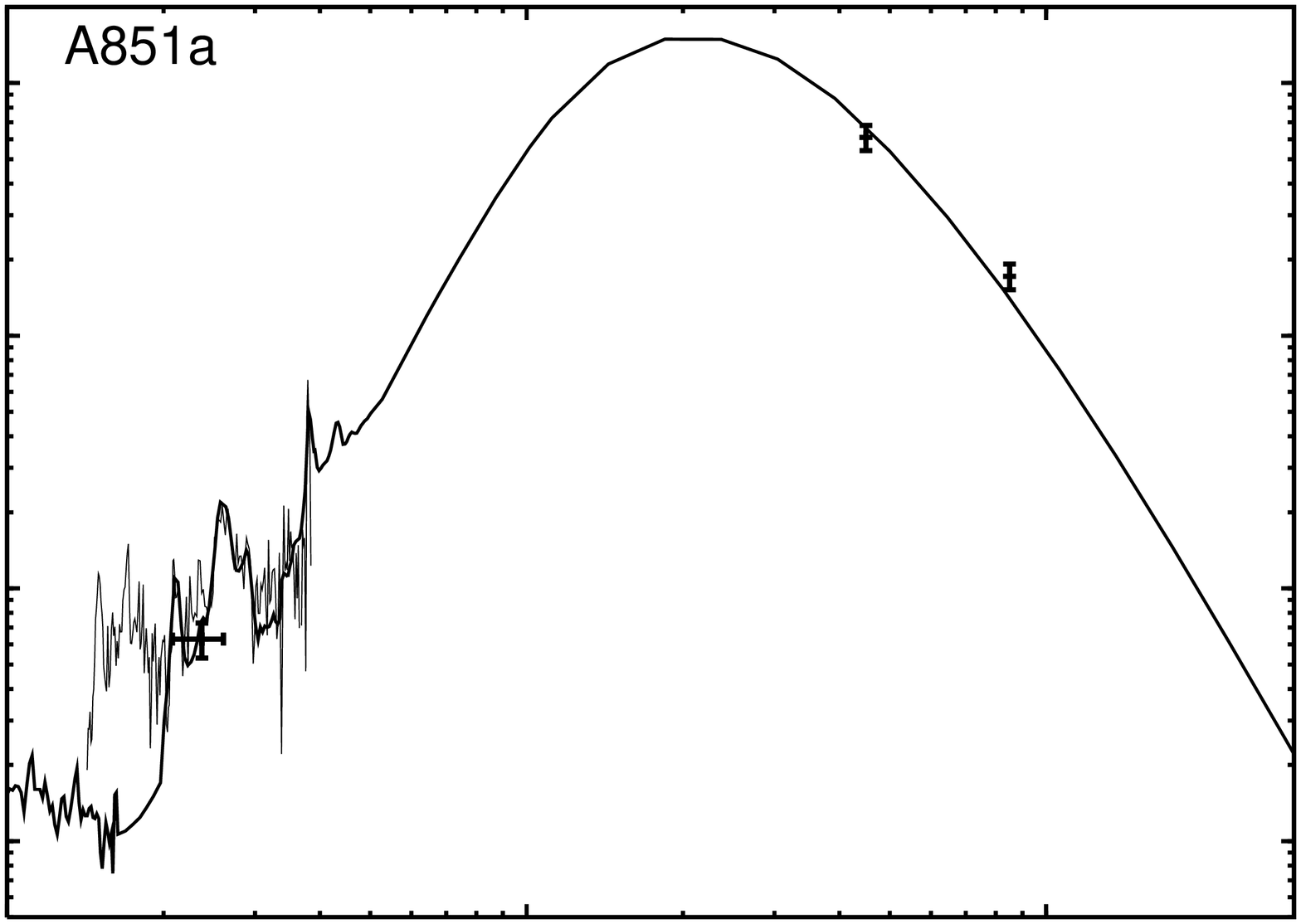}
\includegraphics[width=1.85in,angle=270]{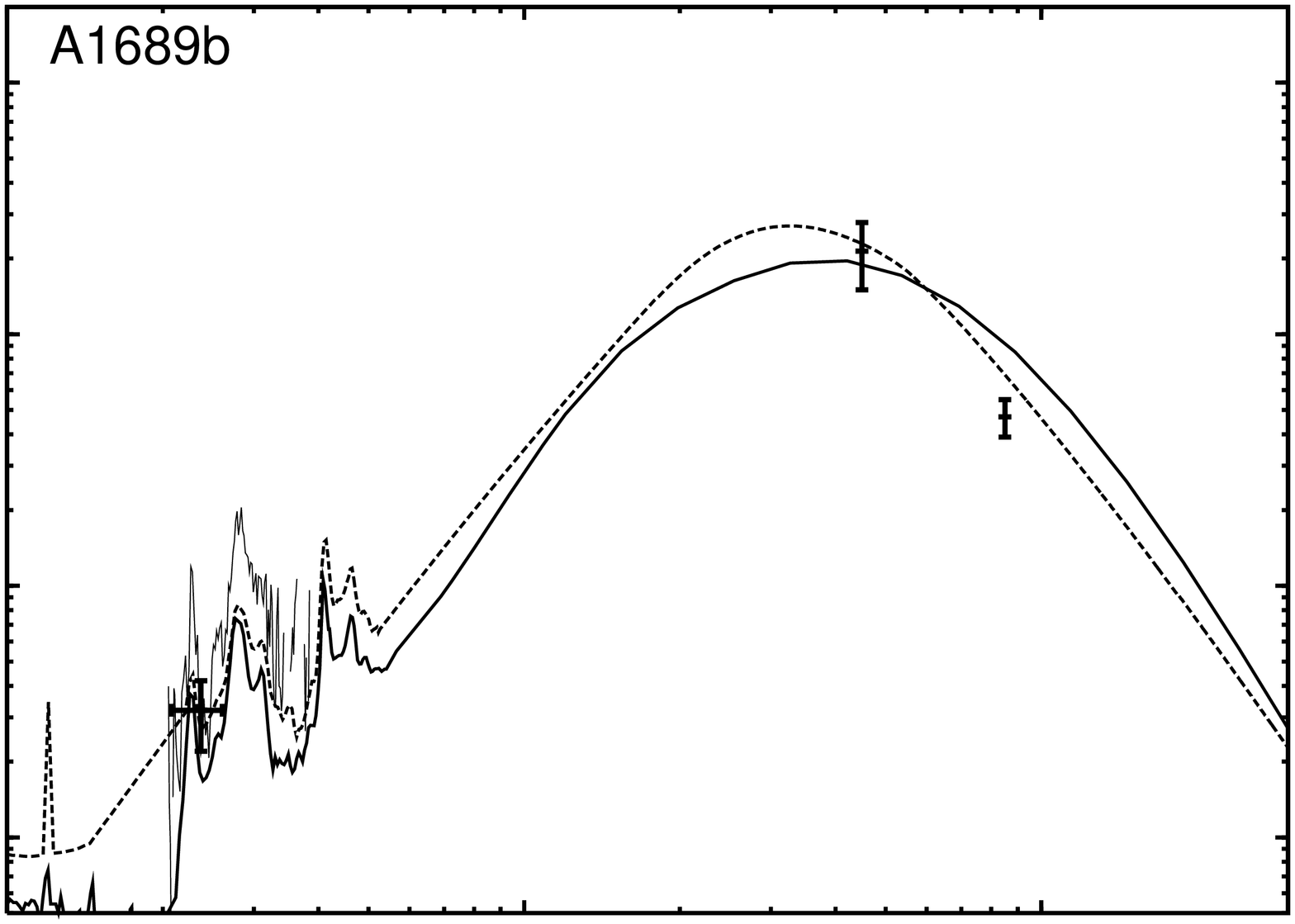}
\includegraphics[width=1.85in,angle=270]{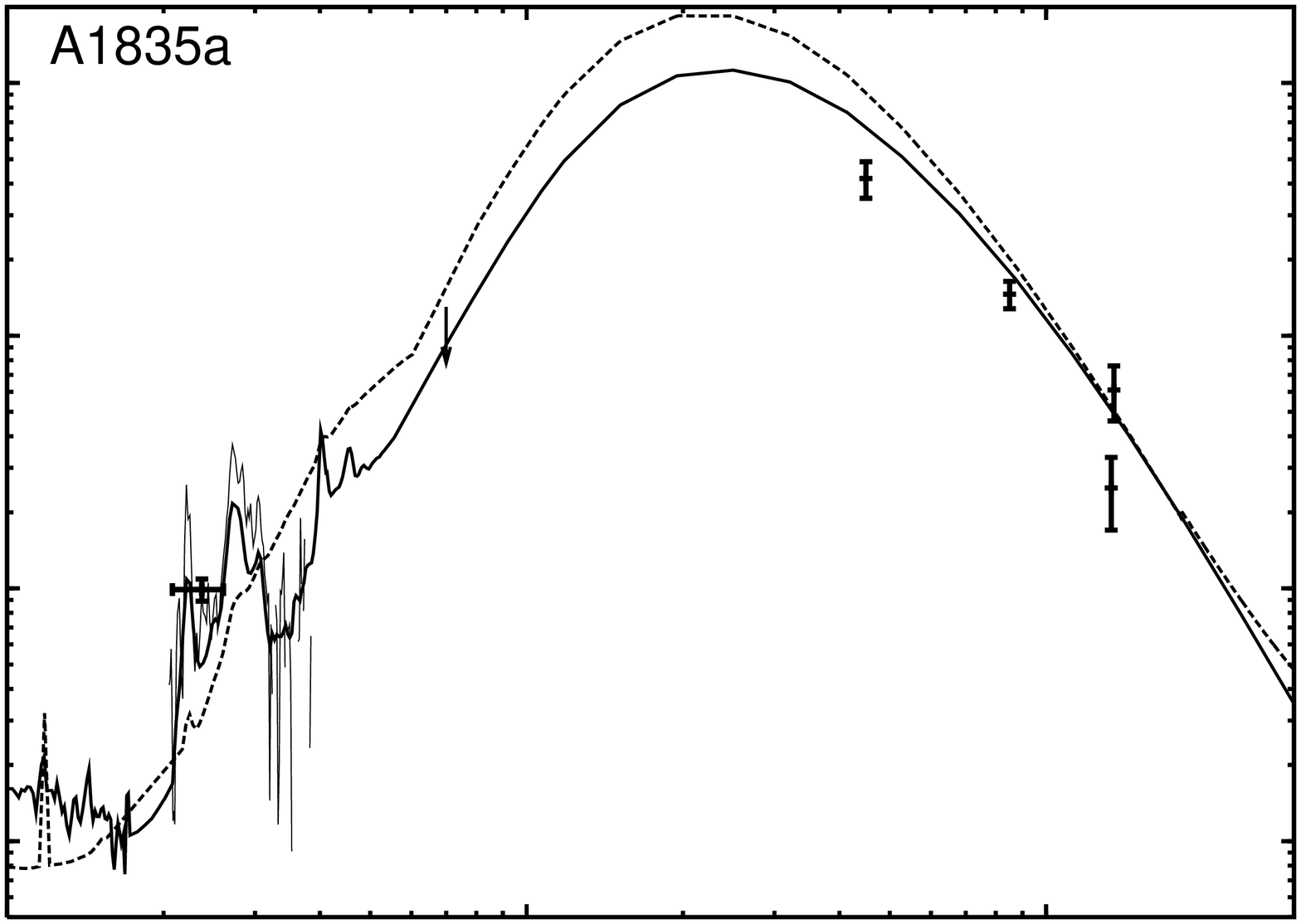}
\includegraphics[width=1.85in,angle=270]{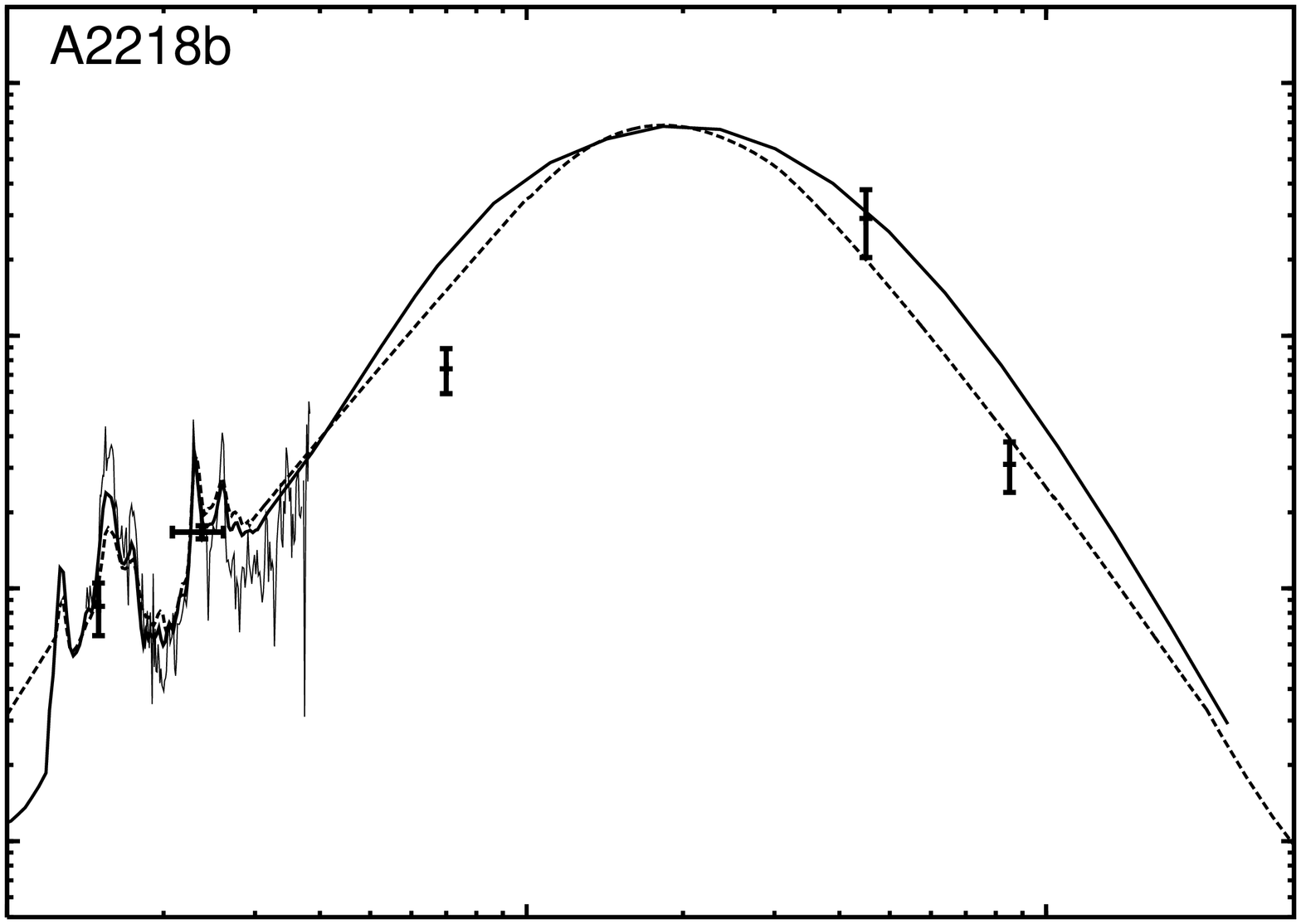}
\includegraphics[width=1.85in,angle=270]{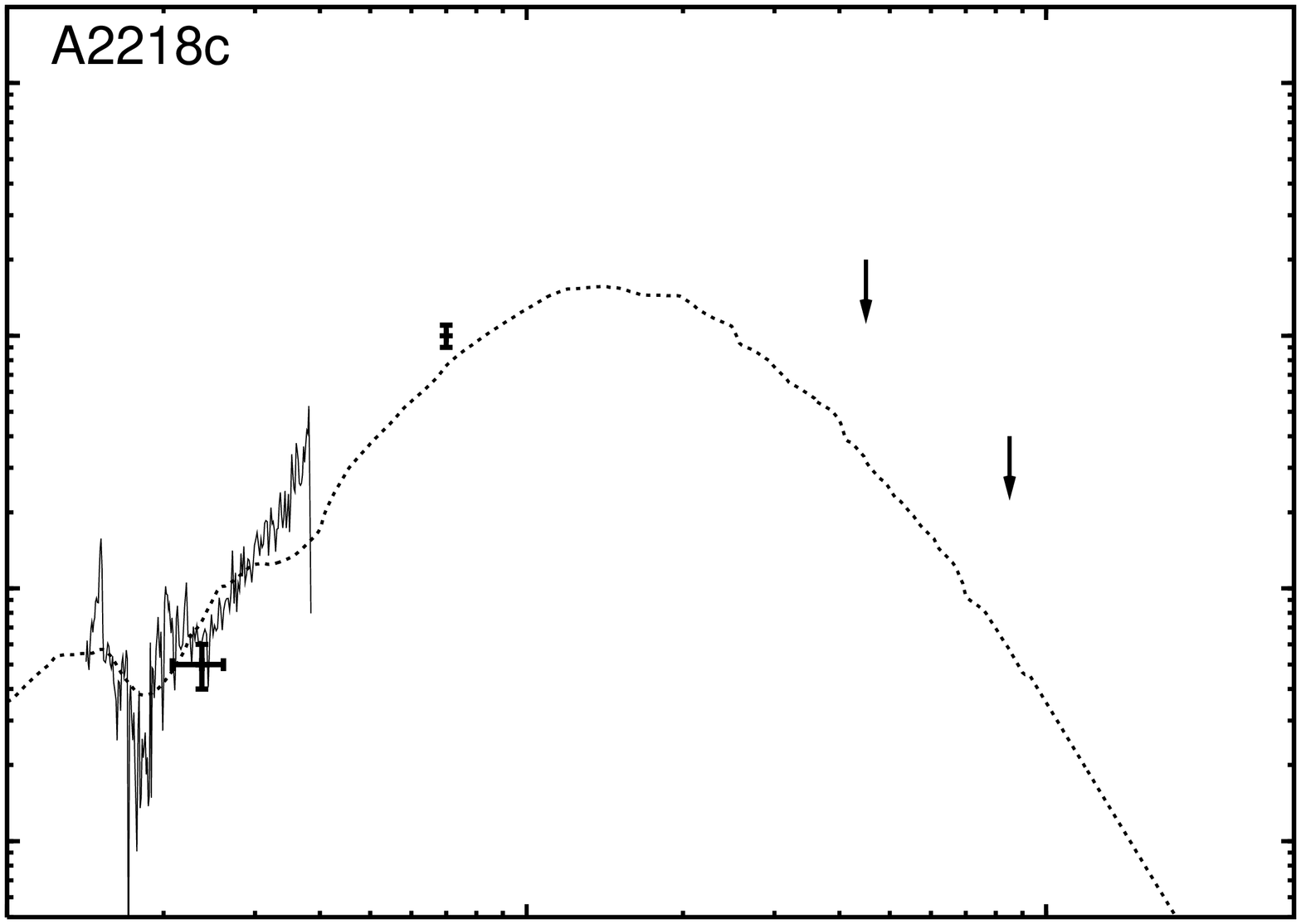}
\includegraphics[width=1.85in,angle=270]{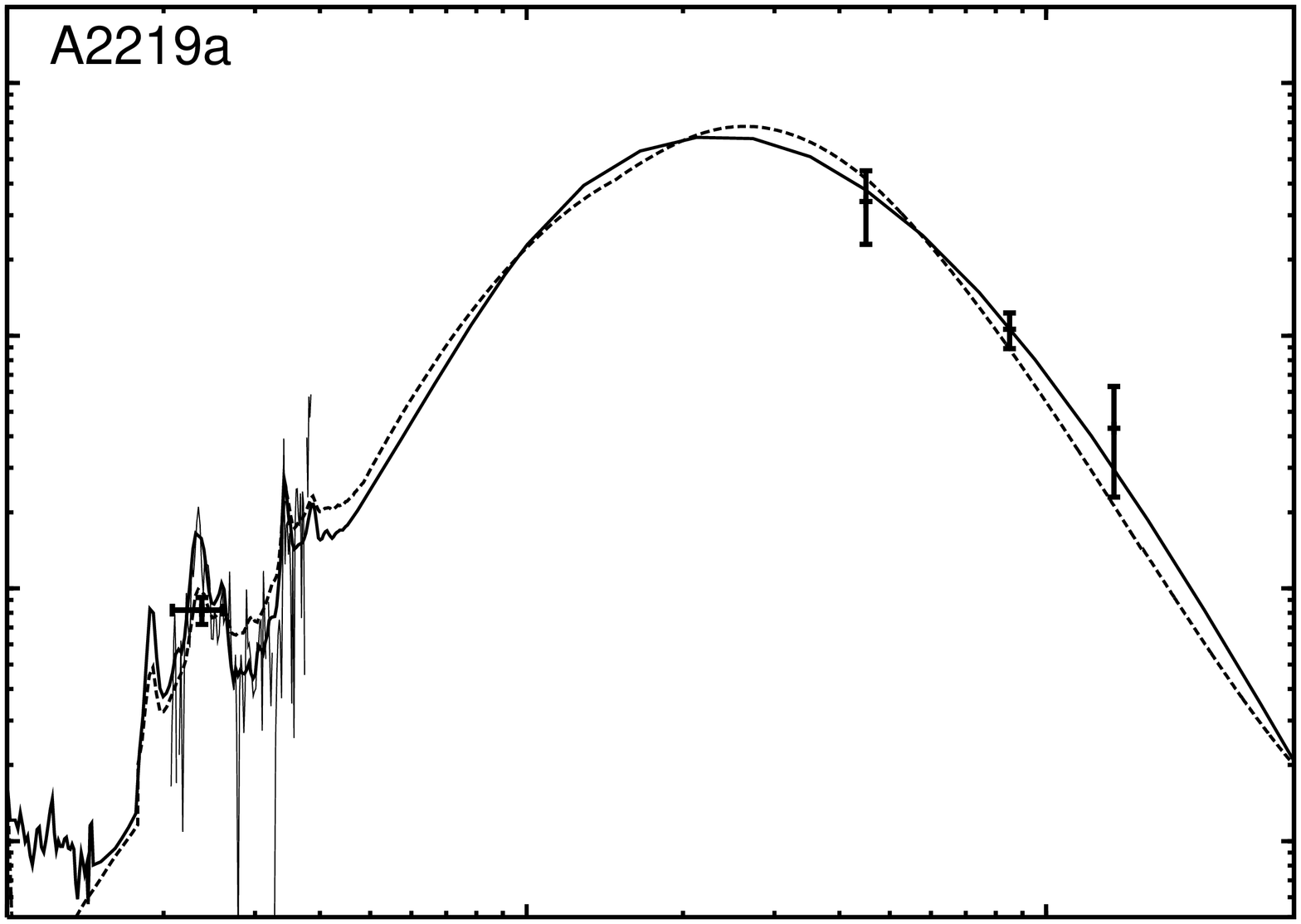}
\includegraphics[width=1.85in,angle=270]{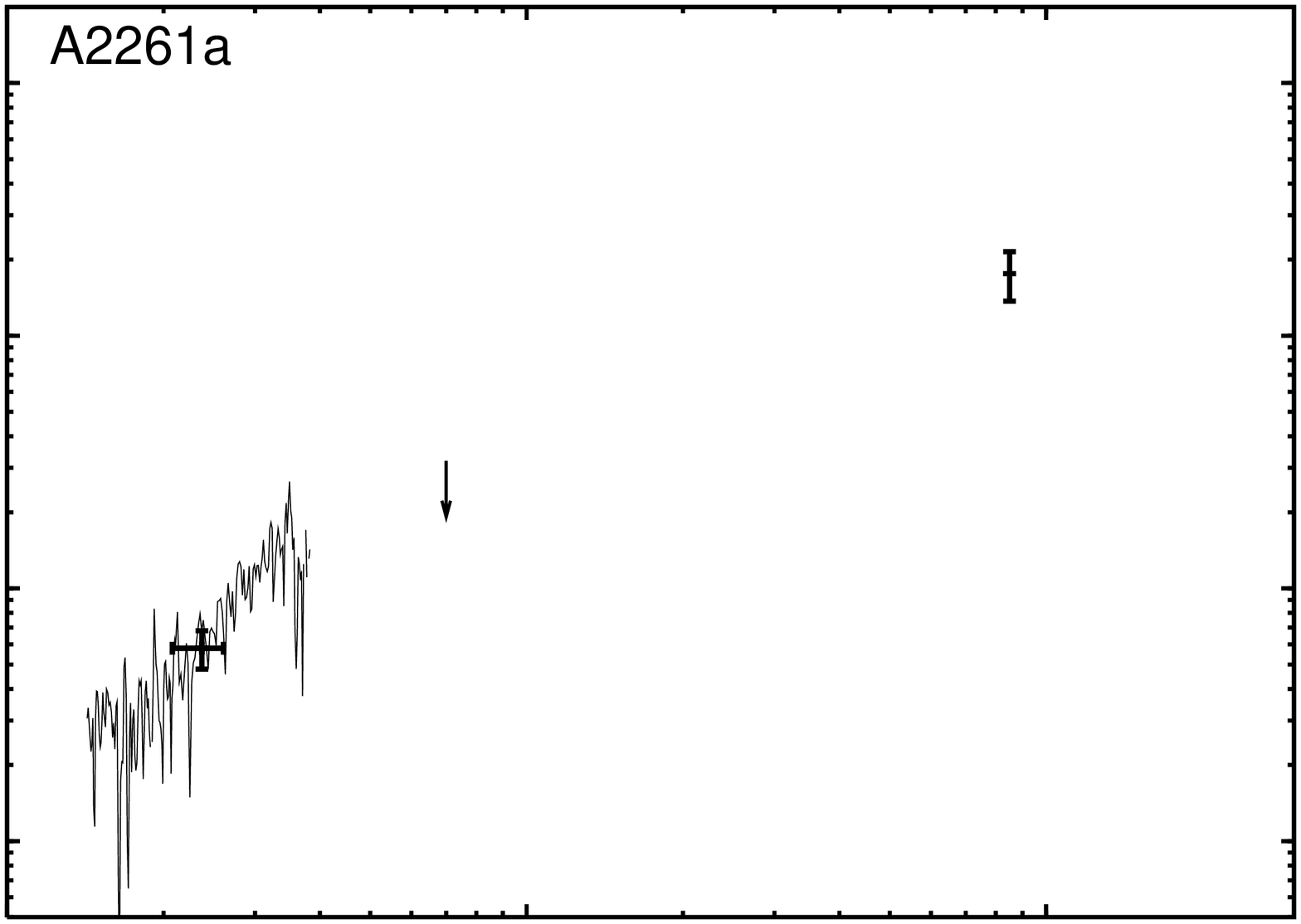}
\includegraphics[width=2.0in,angle=270]{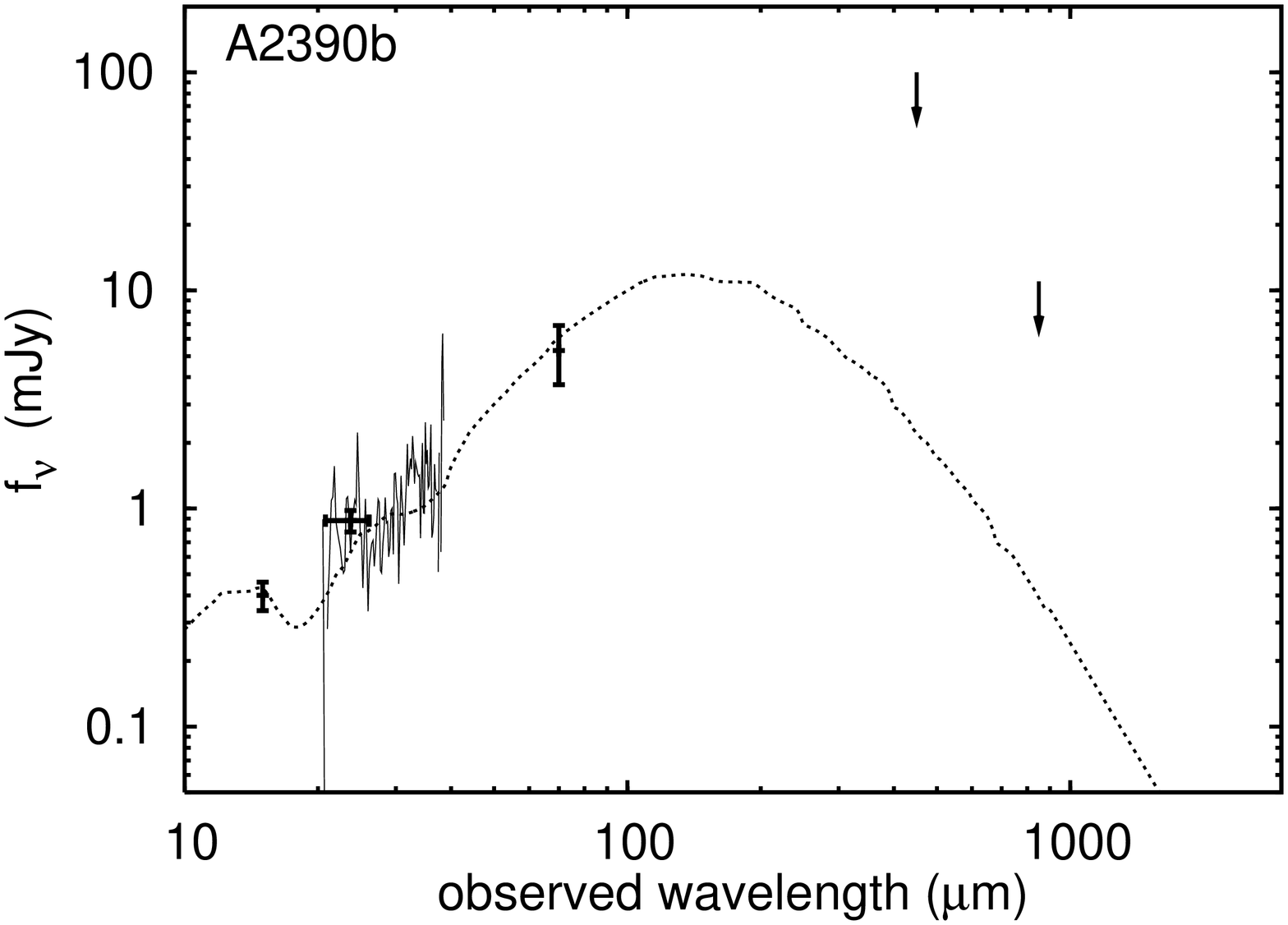}
\includegraphics[width=2.0in,angle=270]{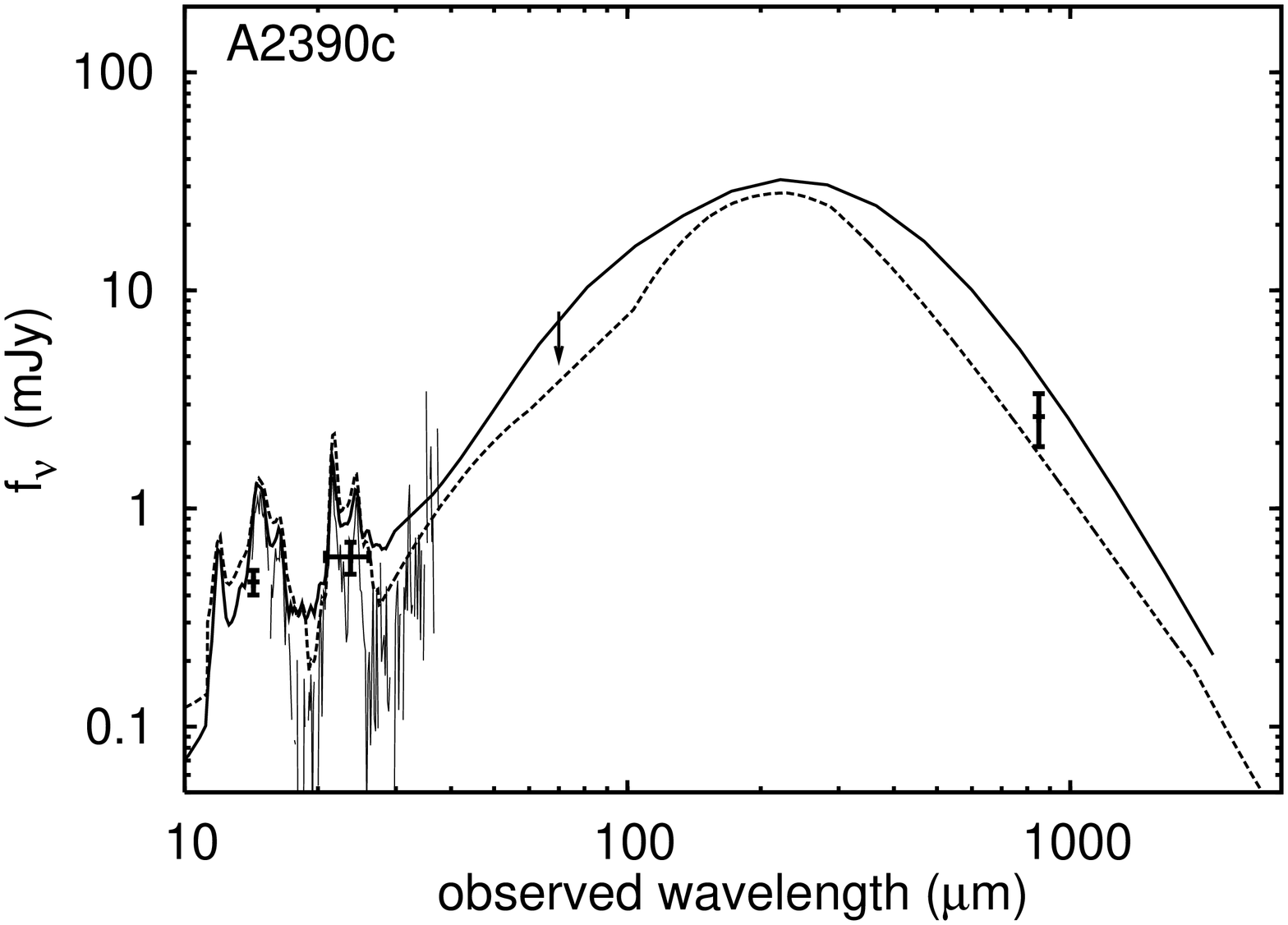}
\figcaption{Photometry and L(TIR) fits, continued.}
\label{fig:other_phot}
\end{figure}
%more notes to add:  
%% No CE fit is plotted for A1835a because there were no acceptable fits.
%%

%%%%%%%%%%%%%%%%%%%%%
\begin{figure}
\figurenum{7}
\includegraphics[width=5in,angle=0]{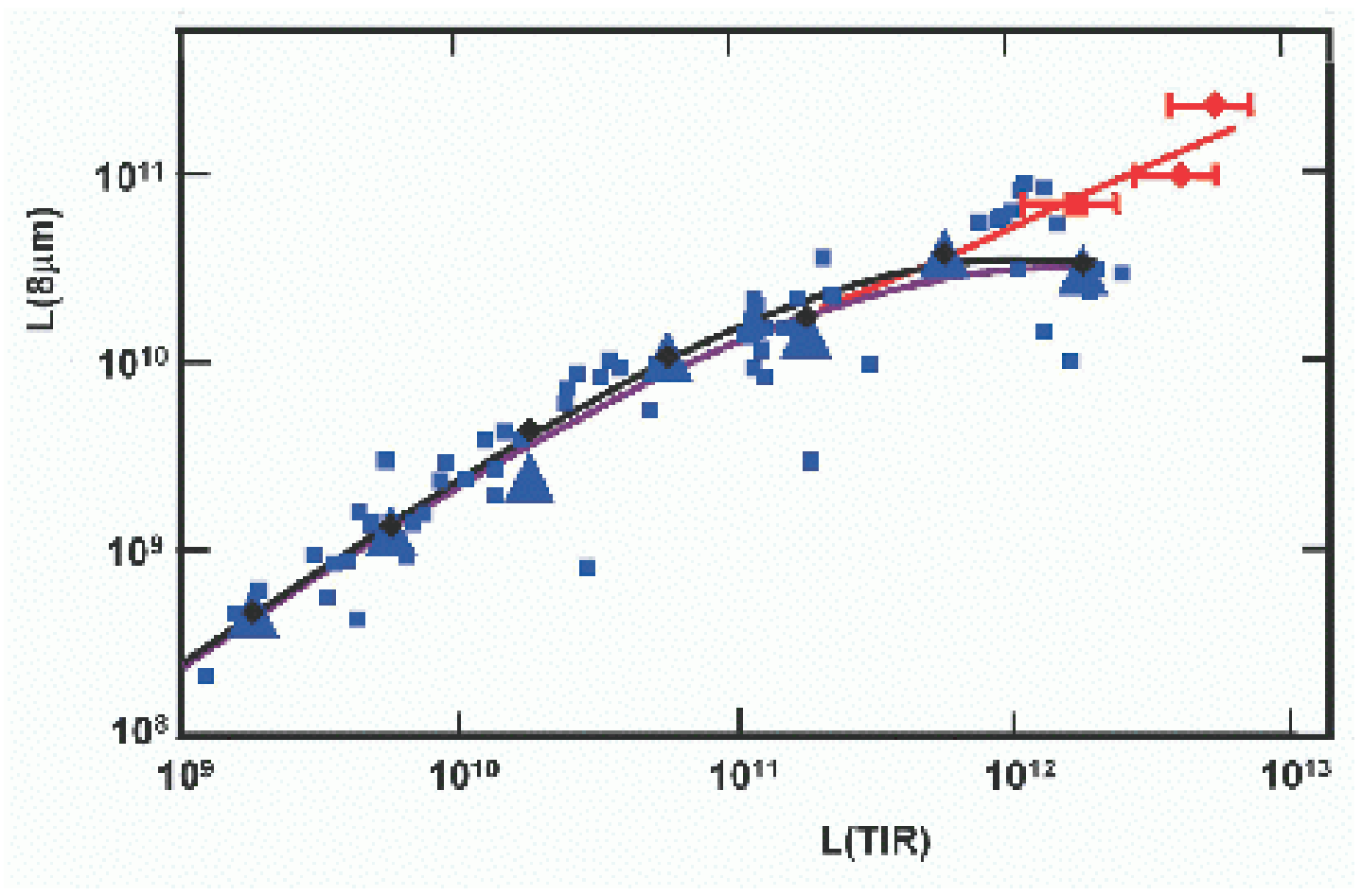}  %{lumvslumb.eps}
\figcaption{
Comparison of L(8$\mu$m) and L(TIR) locally and for stacked
measurements at $z \sim 2$. The individual local galaxies are
indicated as small blue squares, while the values averaged over L(TIR)
intervals of 0.5 dex are shown as triangles. 
The sizes of triangles indicate the uncertanties in the average values 
for the bins, estimated from the scatter in the values.
A fit to the averages is shown as a blue line. 
The black diamonds and line show a similar fit after rejecting high and 
low outliers in the luminosity bins (see text). 
The stacked results at $z \sim 2$ are
shown as a red square \citep{daddi05} and  two red diamonds 
\citep{papovich07}.  The errors in these results have been consolidated into
errors in L(TIR), since the conversion to this parameter is the
dominant uncertainty. The error bars show uncertainties by a factor of
1.5 for \citet{daddi05} and 1.4 for \citet{papovich07}.   The
red line is a fit to the $z \sim 2$ stacked points, constrained to
agree with the local fit at $10^{11}$~\Lsun.
}
\label{fig:TIRL8}
\end{figure}

\begin{figure}
\figurenum{8}
\includegraphics[width=5in,angle=0]{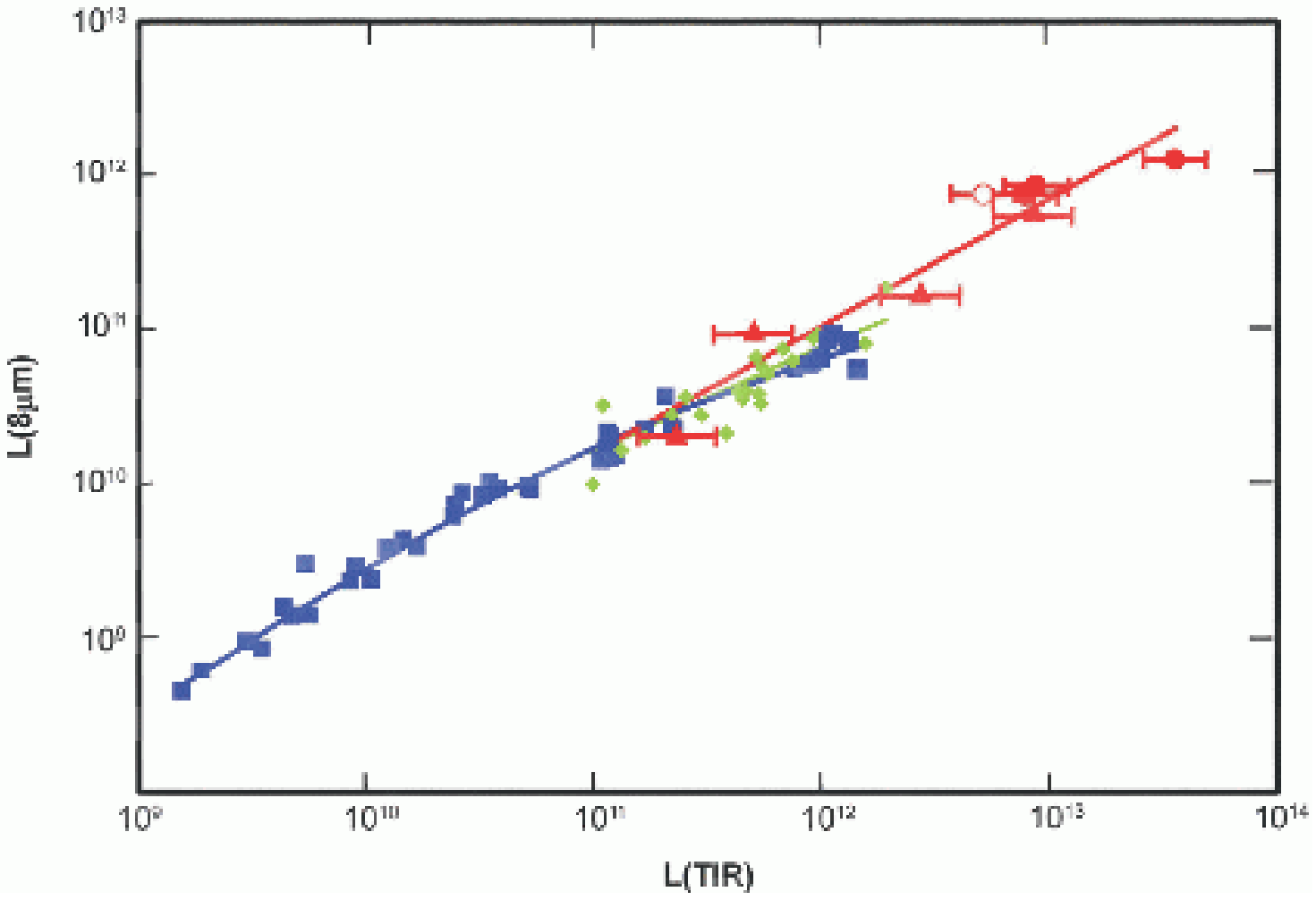}
\figcaption{
Comparison of L(8$\mu$m) and L(TIR) locally and for galaxies at z
$\sim$ 2 measured individually. The individual local galaxies with
above average L(8$\mu$m)/L(TIR) are indicated as blue squares. The
blue line is a fit to them. The high-z galaxies from \citet{yan07}
are shown as red circles, filled for those detected at 1.4GHz and open
for the undetected example. The filled triangles are the lensed
galaxies from this work. The errors have been consolidated into
L(TIR). They are shown as a factor of 1.4 for the \citet{yan07}
galaxies, equivalent to the ratio of flux densities at 60$\mu$m to
1.4GHz varying from 100 to 200. For the lensed galaxies, we show
errors by a factor of 1.5, based on the range of luminosities
indicated by our template fits. The red line is a fit to the z $\sim$
2 points, constrained to agree with the local fit at $10^{11}$~\Lsun. 
The green points are individual galaxies at $z \sim 0.85$,
from \citet{delphine06} and the green line is a fit to those
points constrained to agree with the local fit at $10^{11}$~\Lsun.
}
\label{fig:newTIR}
\end{figure}

%%%%%%%%%%%%%%%%%%%%%
\begin{figure}
\figurenum{9}
\includegraphics[height=6in,angle=270]{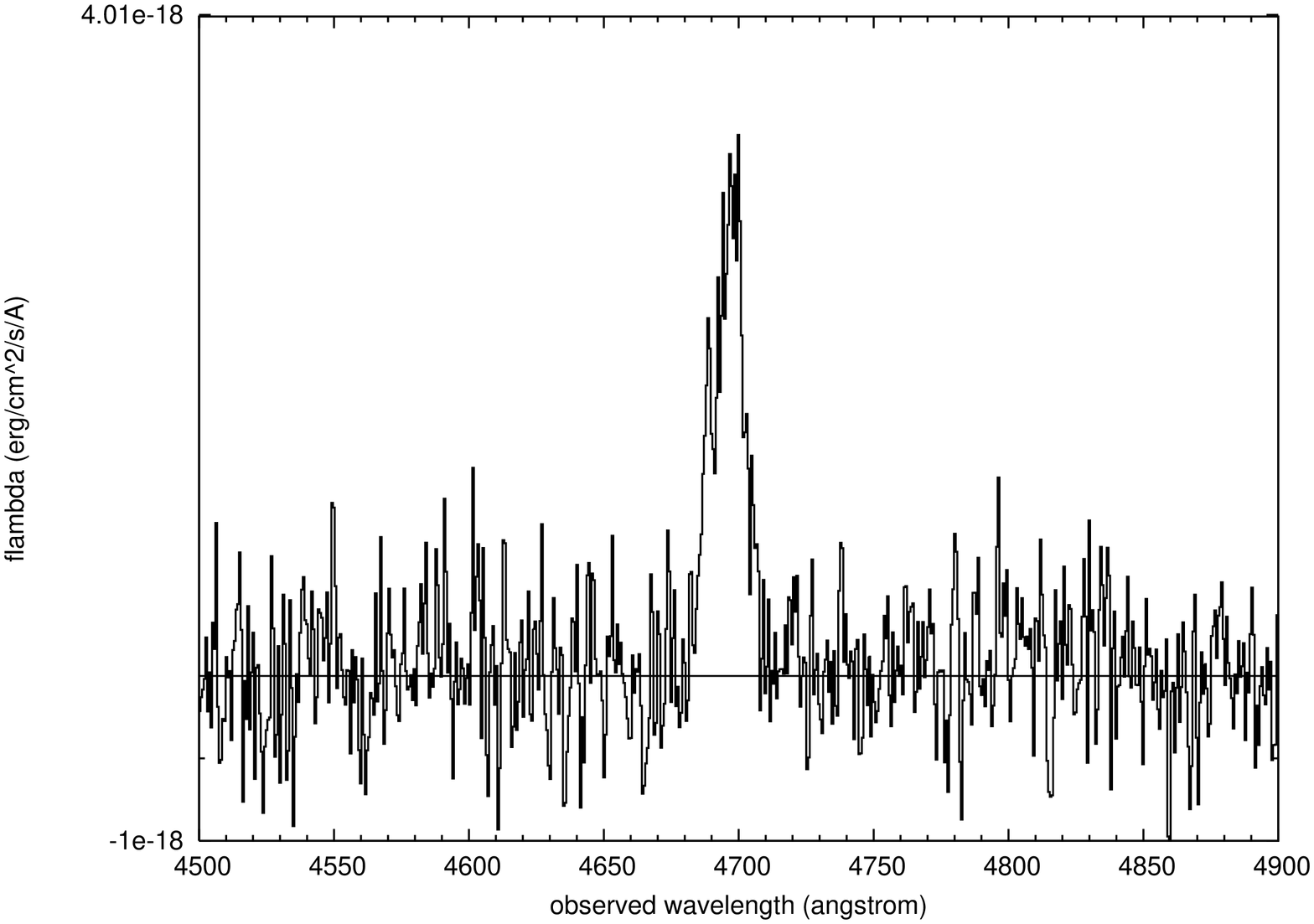}
%{/WORK/Clusters/IRS_spectra_lensed/Opt_spec/johan_spec.ps}
\figcaption{Keck/LRIS spectrum of  A2390a.  The 
strong feature at $\lambda = 4690$~\AA\ is presumably 
Lyman $\alpha$ at $z=2.858$.}
\label{fig:johan_spec}
\end{figure}

%%%%%%%%%%%%%%%%%%%%%%%%%%%%%%
\end{document}